\documentclass[aps,pra,twocolumn,longbibliography]{revtex4-1}

\usepackage{graphicx}
\usepackage{amsfonts}
\usepackage{amsmath}
\usepackage{amssymb}
\usepackage{color}
\usepackage{bbold}
\usepackage{bm}
\usepackage{bbm}
\usepackage{enumerate}
\usepackage{color}
\usepackage{mathtools}
\usepackage{array}
\usepackage[colorlinks,bookmarks=false,citecolor=blue,linkcolor=red,urlcolor=blue]{hyperref}
\usepackage[squaren]{SIunits}

\newcommand{\eg}{\textit{e.g. }}
\newcommand{\ie}{\textit{i.e. }}

\begin{document}

\title{Probing quantum correlations in many-body systems: a review of scalable methods}
\date{\today}

\author{Irénée Frérot\footnote{irenee.frerot@neel.cnrs.fr}}
\affiliation{Institut Néel, CNRS, Grenoble, France}

\author{Matteo Fadel\footnote{fadelm@phys.ethz.ch}}
\affiliation{Department of Physics, ETH Z\"{u}rich, Switzerland}

\author{Maciej Lewenstein\footnote{maciej.lewenstein@icfo.eu}}
\affiliation{ICFO-Institut de Ciencies Fotoniques, The Barcelona Institute of Science and Technology, Castelldefels (Barcelona) 08860, Spain.}
\affiliation{ICREA, Pg.~Lluís Companys 23, 08010 Barcelona, Spain.}

\begin{abstract}
    We review methods that allow one to detect and characterise quantum correlations in many-body systems, with a special focus on approaches which are scalable. Namely, those applicable to systems with many degrees of freedom, without requiring a number of measurements or computational resources to analyze the data that scale exponentially with the system size. We begin with introducing the concepts of quantum entanglement, Einstein-Podolsky-Rosen steering, and Bell nonlocality in the bipartite scenario, to then present their multipartite generalisation. We review recent progress on characterizing these quantum correlations from partial information on the system state, such as through data-driven methods or witnesses based on low-order moments of collective observables. We then review state-of-the-art experiments that demonstrated the preparation, manipulation and detection of highly-entangled many-body systems. For each platform (e.g. atoms, ions, photons, superconducting circuits) we illustrate the available toolbox for state preparation and measurement, emphasizing the challenges that each system poses.
    To conclude, we present a list of timely open problems in the field.
\end{abstract}
\maketitle

\tableofcontents

\section{Introduction}
In both classical and quantum many-body physics, it is a core issue to characterize statistical correlations among different degrees of freedom. From this point of view, a natural question is to establish fundamental differences between quantum and classical many-body systems. In essence, the phenomenon of quantum entanglement (including its various manifestations such as Bell nonlocality) demonstrates the fundamental limitations of classical physics to explain correlations in composite systems. In addition to establishing a clear separation between classical and quantum systems from the point of view of the correlations they host, probing the manifestations of quantum entanglement in many-body systems sheds light on at least three major problems of basic science:
\begin{itemize}
    \item Is there a quantum-to-classical transition for systems of increasing size and complexity? Are multipartite quantum correlations fundamentally different from their bipartite counterparts? 
    \item Can quantum correlations be a resource for tasks inaccessible by classical means? What is the role of many-body quantum correlations in quantum computing, simulation, and metrology?
     \item Do quantum correlations play a role in determining collective properties of many-body systems? Can they be associated to new phases of matter or phase transitions, and how do they affect systems both at and out of thermal equilibrium?
\end{itemize}
The answers to all the questions above are, in a broad sense, affirmative, but highly nontrivial.

\subsection{Contemporary approaches to information processing}
Over the last decades, research on information processing by quantum systems led to a change of paradigm in information science.\\

\noindent{\it Classical information theory.--} As originally developed by Shannon and coworkers, information processing by classical systems is {\it device independent} \cite{Shannon,Cover,McKay}.  As an example, the rules governing the processing of elementary units of information (bits), are the same for modern supercomputers as for Charles Babbage's nineteenth century mechanical machines \cite{Babbage-wiki,Gleick}. In essence, the only differences lie in the speed of information processing and the size of the memory. In practice, though, the above statement can be very misleading. As pointed out by Landauer, every information processing is a physical process that must be associated with energetic costs. Landauer's principle \cite{Landauer,Bennett-Landauer} states that ``any logically irreversible manipulation of information, such as the erasure of a bit or the merging of two computation paths, must be accompanied by a corresponding entropy increase in non-information-bearing degrees of freedom of the information-processing apparatus or its environment". Concretely, if an observer loses information about a physical system, heat is generated and the observer loses the ability to extract useful work from that system. While the original formulation of Landauer's principle is about information loss, in practice any classical information processing leads to heating. A possible way to understand this basic fact is to consider the change of a bit state being erasing of the previous state, followed by generating a new one. As a matter of fact, heating is one of the main technological challenges of contemporary supercomputers, although heat is mostly generated by the resistance of elements carrying control currents, and exceed by orders of magnitude the fundamental Landauer's bound.\\

\noindent\textit{Quantum information theory.--} In contrast to classical information theory, information processing by quantum systems is based of the rules of quantum mechanics, and thus is in principle strongly {\it device dependent}, pointed out in seminal works in the field (cf.~refs.~\cite{Deutsch?,Bennett?}, see also \cite{Nielsen,Wilde}). The above statement has been mitigated more recently. Stimulated by novel types of attacks proposed by Makarov and coworkers in 2005 against quantum cryptography devices produced by IDQuantique \cite{Makarov,Makarov1}, a new paradigm for {\it device-independent quantum information processing} (DIQIP) was developed \cite{DIQIP,diqip2}. Such a framework is, in essence, based only on the raw input/output data as provided by a ``black box'' device, and it aims at establishing quantum information processing protocols which do not make assumptions on the underlying device. In the recent research on quantum information science, DIQIP approaches (or partial DIQIP approaches, based on data and also various levels of working assumptions about the device), have dominated the literature on detection, characterization, validation and verification of the {\it quantumness} of the said protocols. In  particular, a large number of methods, especially suited to characterize the quantumness of many-body systems in a device-independent perspective, have been developed. A key issue is the possibility to perform such certification tasks in a scalable way, namely, with a computational cost which does not grows exponentially with the system size, as would be naively expected from the exponential growth of Hilbert space dimension for quantum many-body systems. As certification should be based on experimental data as produced in existing or near-term devices, such recent approaches have proposed to focus on very partial information, such as coarse-grained features of the statistical correlations among the local degrees of freedom involving only low-order moments (mainly first and second moments) of local observables. The present review focuses mostly on such methods.

\subsection{The quest for quantum advantage}
We are currently witnessing a spectacular development of quantum  technologies (QT) \cite{roadmapEU,RecentUSA}, which aim at exploiting genuine quantum effects such as quantum entanglement and Bell nonlocality to achieve information-processing tasks with performances which cannot be matched by classical devices. Obviously, if one looks for a quantum advantage of quantum over classical devices, one must manipulate quantum systems that exhibit {\it genuine quantumness}, such as \eg quantum entanglement. In practice, thus, any reliable QT device must allow for the detection, characterization, validation and verification of various facets of its quantumness as a resource. This includes quantum coherence \cite{PlenioRMP}, quantum entanglement \cite{horodecki_RMP_2009} and Bell correlation \cite{brunneretal_RMP_2014,Scarani-book}. Schematically, QT are based on the following pillars.\\ 

\noindent{\it Quantum Communications.--} Quantum communication is perhaps the area of QT which is developed at most, with several companies already offering commercial quantum cryptography systems and quantum random number generators. At a very basic level, in quantum communication we deal with bipartite systems (traditionally called Alice and Bob), and the goal is to ensure rapid communication with the largest possible privacy and integrity of the transmitted information. In cryptographic schemes as originally envisioned by Ekert \cite{Ekert}, the parties ensure privacy by checking the violation of a Bell inequality using a subset of the transmitted data.  The send-and-measure scheme as proposed by Bennett and Brassard \cite{BB84} requires a certain degree of  entanglement that can be tested on the data \cite{Curty1,Curty2}. The situation is more complex when one deals with quantum networks \cite{perc,perc1,perc2,perc3}, or so-called quantum internet \cite{qinternet}, which are inherent many-body problems. As such, they i) may exhibit genuine many-body entanglement and Bell nonlocality; ii) are much more difficult to characterize and probe. Similar situation occurs with quantum random number generators (QRNGs). There, one can certify randomness using Bell correlations \cite{Tony-Nature}, but in practice thiss method is slow. Thus in practice, commonly used QRNGs cannot be truly certified, but rather can be benchmarked by applying long sequences of random numbers in Monte-Carlo studies of classical or quantum many-body problems \cite{quside-paper}.\\

\noindent{\it Quantum Metrology and Sensing.--} The field of quantum sensing has witnessed exquisite progress and successes over the recent years. In this context, genuine quantum correlations are used in the first place to improve precision and accuracy of measurements \cite{GLM11,pezzeRMP}. For a review of quantum metrology with photonic platforms see ref.~\cite{polino2020photonic,gobel2015quantum,barbieri2022optical}, and with atomic ensembles ref.~\cite{pezzeRMP}. In quantum metrology, the most interesting application of quantum entanglement is with massively entangled many-body states (such as spin-squeezed and Dicke states \cite{pezzeRMP}, as well as Schr\"odinger cat states \cite{Paris}). However, the detection and characterization of highly-entangled multipartite states such as Schr\"odinger cat states is a formidable challenge \cite{cirac1998quantum}, requiring for large cats extremely demanding resources \cite{skotiniotis2017macroscopic}. Novel methods based on so-called {\it shadow tomography} \cite{Preskill-shadow,sabrina,randomized_review} might become particularly useful to characterize complex many-body entangled states useful for quantum metrology.\\

\noindent{\it Quantum Computing and Simulation.--} Contemporary QT face challenges in achieving fault tolerant quantum computing with error correction, and focus instead on various shades of quantum simulation devices (Noisy Intermediate Scale Quantum, NISQ) \cite{Preskill18}, analogue and digital quantum simulators \cite{GAN14} and so-called quantum annealers \cite{FFG01}. There is a clear need and quest for systems that, without necessarily simulating quantum dynamics of some physical systems, can generate massively-entangled and superposition states in a controlled and robust way. This should in particular include the control of decoherence, enabling the use of these states for quantum communications \cite{GT07}, quantum metrology \cite{GLM11}, sensing and diagnostics \cite{DRC17} (e.g. to precisely measure phase shifts of light fields, or to diagnose quantum materials). In a recent essay \cite{LNP1000}, a vision of the future of quantum simulators over the coming decades is presented. This essay also contains a brief review of the current status of quantum computing, both in the scope of future fault tolerant quantum computers as well as NISQ devices \cite{Alba}. NISQ devices and quantum simulators are also already accessible commercially, and many companies are working on scaling up such machines towards industrial applications. It is interesting to notice that several recent claims of quantum supremacy \cite{Google}, have been challenged by contemporary classical simulations using tensor networks \cite{Pan1,Pan2}. Furthermore, the prospects for exponential  quantum supremacy in quantum chemistry have been questioned \cite{Garnet}. 

In general, the applications of quantum simulators of NISQ devices are {\it per se} devoted to many-body systems. Developing methods for the detection, characterization, validation and verification of quantum correlations in such systems is hence of primary importance for the whole field. As already emphasized, such methods should be scalable, be based preferentially on raw data (namely, on results of experimental measurements), and ideally they should be device independent. Matching all such expectations poses important challenges, and recent methods in this direction are reviewed here.

\subsection{Two-fold way}
The challenge of entanglement detection and characterization in many-body systems has been attacked in the recent decade with mainly two complementary approaches.\\

\noindent{\it Methods based on random unitaries.--} One of the most fundamental quantities that characterizes entanglement in many-body systems is the entanglement entropy \cite{eisert_etal_2010}, \ie the von Neumann or Rényi entropy of the reduced density matrix of a subsystem. Such quantities are very difficult to measure in experiments, and their interpretation as entanglement measures requires to assume that the global state is pure, a condition which is never exactly matched in an experiment. Rényi entropy of order two ($S_2$) was measured in several experiments involving the preparation of two identical copies of the system \cite{Islam2015,KaufmanSci16}, involving up to eight qubits in the most recent experiments \cite{Lukin_2019,Bluvstein_2022}. As such an approach is very hard to scale up, alternative methods based on applying random unitaries before measuring in the computational basis have been developed \cite{elbenetal2018,Brydges_2019,benoit3,vermerschetal2019}. They allow one not only to measure $S_2$ on a single copy of the system, but also to extract many other crucial properties to characterize quantum many-body systems, such as measuring topological invariants \cite{benoit6} (see ref.~\cite{randomized_review} for a recent review). Sampling over random unitaries is also a central tool in so-called shadow tomography protocols \cite{Preskill-shadow,sabrina}. In the conventional quantum tomography of $N$ qubits, exponentially-many measurements are required in order to reconstruct the density-matrix. If one, however, is only interested in estimating the average of a finite number of observables, then the number of necessary measurements can be reduced to be linear in the number of qubits. Random-unitaries methods are remarkably powerful, yet they come with intrinsic limitations:
\begin{enumerate}
    \item Experimentally, they require an exquisite control over the individual qubits, to both manipulate and measure them. Such degree of control is not available to all experimental platforms relevant to quantum metrology and quantum simulation.
    \item The estimation of entanglement entropies is very suitable for at most a few tens of qubits, but is hardly scalable to very large systems of several hundreds or thousands of individual components.
\end{enumerate}

\noindent{\it Methods based on low order moments.--} Complementary approaches based on partial, and potentially very coarse-grain, information on the underlying quantum many-body state, have therefore also been developed. In such approaches, which are the main focus of this review, one typically tries to construct entanglement witnesses, Bell inequalities, quantum coherence measures, and more generally quantumness tests based on experimentally available data. These often correspond to measurements of low order (often first and second) moments of some local observables, or from the knowledge of a few average values of collective many-body observables. Frequently such approaches leverage on symmetries, such as permutation or translation symmetry, and thus focus on averages and fluctuations of global quantities such as collective spin components, which are the sums of local quantities (spins of the system's constituents: atoms, molecules, etc.), or various components of the structure factors. Thus, the corresponding entanglement criteria or Bell inequalities may be viewed as generalizations of so-called spin squeezing inequalities \cite{wineland1992,SorensenNat01}, which were originally introduced in the context of quantum metrology.

\begin{figure}
    \centering
    \includegraphics[width=0.8\columnwidth]{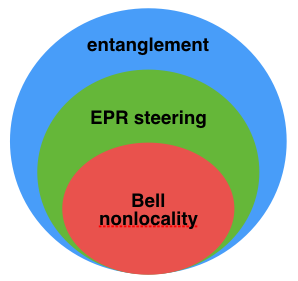}
    \caption{\textbf{Hierarchy of quantum correlations.} Three main types of correlations can be identified within the framework of quantum mechanics: entanglement, EPR steering, and Bell nonlocality. These represent inequivalent resources for quantum information tasks, and form a hierarchy of correlations, in the sense that entanglement is necessary but insufficient to have EPR steering, and EPR steering is necessary but insufficient to have Bell nonlocality.}
    \label{fig:hierarchy}
\end{figure}

\subsection{Guide through existing reviews}
Several excellent reviews on various aspects of quantum correlations are available in the literature. The reader might consult a review on the concept of quantum entanglement from a quantum information theory perspective \cite{horodecki_RMP_2009}, a review discussing recent developments on the concept Bell nonlocality \cite{brunneretal_RMP_2014}, or on Einstein-Podolsky-Rosen (EPR) steering \cite{reidetal_RMP_2009,uola_etal_RMP_2020}. As illustrated in Fig.~\ref{fig:hierarchy} and pedagogically introduced below, quantum entanglement, Bell nonlocality and EPR steering are inequivalent manifestations of quantum nonlocality in a broad sense. 

Entanglement witnesses are reviewed in refs.~\cite{GuhneT_2009} and \cite{chruscinskiS_2014}. A zoo of concepts of quantum correlations weaker than entanglement is reviewed in ref.~\cite{AdessoBC_2016}, including the concept quantum discord which is also independently reviewed in refs.~\cite{modi_etal_RMP_2012,bera_etal_2017}. 

Further fundamental manifestations of quantum nonlocality, such as the very strong form of device-independent certification of quantum devices called self-testing \cite{Supic2020selftestingof}, or recent developments on Bell nonlocality in networks \cite{Tavakoli_2022}, have also been recently reviewed.

Reviews focusing on quantum entanglement in many-body systems have mostly considered bipartite entanglement entropies in ground states. This includes ref.~\cite{eisert_etal_2010} and ref.~\cite{laflorencie_2016}. A review on other aspects of quantum correlations in many-body systems can be found in ref.~\cite{deChiaraS2018}. Quantum aspects of macroscopic states are reviewed in ref.~\cite{frowis_etal_RPM_2018}. Entanglement quantification and certification are reviewed in ref.~\cite{friietal_review_2018}, and its role as a resource for quantum metrology is reviewed in ref.~\cite{pezzeRMP}. Finally, a recent review on applications of random unitaries, including probing of bipartite entanglement, is presented in ref.~\cite{randomized_review}.

In contrast to the above-mentioned works, the present review focuses more specifically on those aspects of quantum entanglement, EPR steering and Bell nonlocality which can be assessed in a scalable way from experimental data as collected in a variety of current-day many-body experiments.

\subsection{The plan of the review}
In this review we will present techniques which probe multipartite quantum correlations, with a special emphasis on those methods which can be applied in a scalable way to systems with many degrees of freedom, namely with a cost (in terms of required measurements and computational resources to analyze the data) which do not scale exponentially with the system size.

In Section~\ref{sec2} we will focus on giving a pedagogical introduction on quantum correlations, starting from introducing the bipartite definitions of entanglement, Einstein-Podolsky-Rosen steering, and Bell nonlocality, to then generalize them for the multipartite case. In Section~\ref{sec3} we will review the theoretical developments on tools that allow to detect multipartite quantum correlations in a scalable way, e.g. from data-driven approaches that use limited amount of measurements, or from analytical approaches that consider collective observables. In Section~\ref{sec4} we will give an overview of the experimental setups used to investigate multipartite quantum correlations in physical systems, such as ensembles of atoms, ions, photons, and superconducting qubits. After mentioning the different state preparation and characterization methods, we will give a selection of results on multipartite correlations that have been achieved. Finally, in Section~\ref{sec5} we will give our conclusions on the discussed topics, and present a list of open questions that could be of interest to the community.

\section{Classifying statistical correlations}\label{sec2}

\subsection{Many flavors of quantum correlations: A pedagogical introduction}
\noindent\textit{Microstates and correlations in classical physics.--}
According to classical physics, the physical world can be decomposed into systems which possess well-defined properties (i.e. elements of reality \cite{EPR}), encapsulated into the concept of a microstate. If some of the system's properties are unknown or unaccessible, one can refer to them as local hidden variables \cite{Bell1964}. 

For any two systems denoted as $A$ and $B$, one can then define the microstate for the joint $AB$ system from the pair $(\vec x_A, \vec x_B)$, with $\vec x_{A(B)}$ specifying the microstate of $A(B)$. In classical physics, a microstate is the most complete description of the composite system $AB$. Eventually, if there is a degree of ignorance about the actual microstate, one can introduce a probability distribution $p(\vec x_A, \vec x_B)$, and then describes the joint $AB$ system according to the statistical properties of $p$. 

In quantum theory, the notion of microstate is replaced by a wavevector $|\psi\rangle$. From the classical idea of independent systems, \ie separated in space and described by their own states, one could then conclude that the most complete description of any composite $AB$ system is contained in the pair of wavevectors $(|\psi_A\rangle, |\psi_B\rangle)$, or by a statistical mixture of such pairs. Remarkably, this is not true.\\

\noindent\textit{Quantum entanglement.--} Quantum theory admits the existence of quantum states $|\psi_{AB}\rangle$ for the composite system $AB$, that cannot be expressed by the independent knowledge of the $A$ and $B$ states. This is the concept of \textit{quantum entanglement} \cite{schrodinger1935}, which encapsulates the idea that the quantum state of a composite system can contain more information than the collection of the states of its constituents. 

Formally, a composite system $AB$ is described by a density matrix $\hat \rho_{AB}$. One can then say that $\hat \rho_{AB}$ is \textit{separable} if there exists a probability distribution $p(|\psi_A\rangle, |\psi_B\rangle)$ such that:
\begin{equation}
    \hat \rho_{AB}^\text{sep} = \sum_{|\psi_A\rangle, |\psi_B\rangle} p(|\psi_A\rangle, |\psi_B\rangle) ~ |\psi_A\rangle \langle \psi_A|\otimes |\psi_B\rangle \langle \psi_B| ~.
    \label{eq_rho_AB_sep_intro}
\end{equation}
In words, $\hat \rho_{AB}$ is separable if it can be written as a statistical mixture of product states $|\psi_A\rangle \otimes |\psi_B\rangle$, with weights given by $p(|\psi_A\rangle, |\psi_B\rangle)$. If that is not the case, namely if $\hat \rho_{AB}$ cannot be decomposed in such a way, then the system is said to be in an \textit{entangled state}.\\

\noindent\textit{Minimal example: a pair of spins.--}
As a pedagogical introduction, let us start with considering a pair spin-$1/2$ particles. Concretely, we imagine a situation where the state of this pair is prepared, and then each particle is sent to a different experimenter. The goal of the two experimenters, Alice and Bob, is then to perform some local spin measurements on their particle, and based on the statistical correlations they observe among the outcomes, determine whether the underlying joint quantum state is separable or entangled.

According to quantum theory, the most complete description of a pair of spin-$1/2$ is given by the density matrix $\hat \rho_{AB}$, namely a $4\times 4$ positive-semidefinite hermitian matrix of unit trace. In this simple situation, it is known that $\hat \rho_{AB}$ is separable if and only if its partial transpose is also a positive matrix \cite{horodecki_RMP_2009,GuhneT_2009}. Therefore, in order to determine whether the state of the system they share is separable or not, Alice and Bob could measure the 16 expectation values $\langle \hat A_i \hat B_j \rangle$, with $\hat A_i$ and $\hat B_j$ chosen among $\{\mathbb{1}, \hat X, \hat Y, \hat Z \}$ (namely the identity and the three Pauli matrices for the respective spin). These allow them to reconstruct $\hat \rho_{AB}= (1/4)\sum_{i,j} \langle \hat A_i \hat B_j \rangle \hat A_i \otimes \hat B_j$, and then to compute the eigenvalues of its partial transpose.

Unfortunately, this remarkable result could never be generalized to situations significantly more complex than a pair of spin-$1/2$. Furthermore, it may be the case that Alice and Bob are not able to perform all measurements required to reconstruct completely the quantum state $\hat \rho_{AB} $, namely to perform full state tomography. It is thus of central interest to address the problem of entanglement detection when only partial information on the system is available, such as from limited set of mean values $\langle \hat A_i \hat B_j \rangle$. In particular, this situation is unavoidable in many-body systems (which are the main focus of this review), since full tomography would require a number of measurements scaling exponentially with the number of particles.\\

\begin{figure}
    \centering
    \includegraphics[width=1.\linewidth]{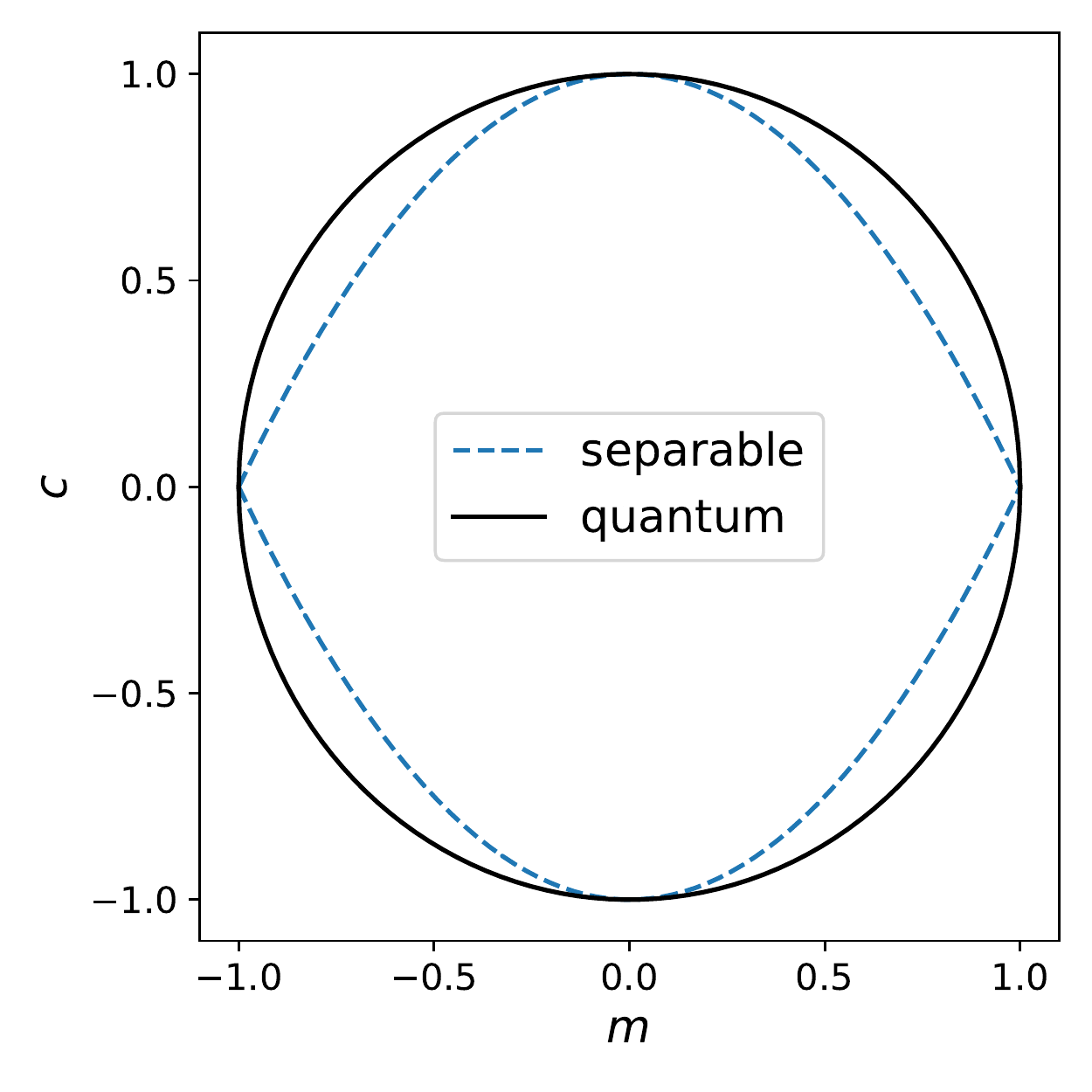}
    \caption{\textbf{Separable versus entangled quantum states.} Considering measurements of $m$ and $c$ as defined in Eqs.~(\ref{eq_def_m_2qubits},\ref{eq_def_c_2qubits}), the black solid circle [resp.~blue dashed line, Eq.~\eqref{eq_wit_2qubits}] delimits the region of values accessible to entangled (resp.~separable) quantum states of two qubits.}
    \label{fig:2qubits_sep}
\end{figure}

\noindent\textit{Entanglement detection from partial information.--}
For the sake of illustrating pedagogically the main concepts behind the detection of quantum entanglement from partial information, let us assume that we only know the mean value
\begin{equation}
   m=\langle \hat Z_A + \hat Z_B \rangle/2
   \label{eq_def_m_2qubits}
\end{equation}
and the correlation
\begin{equation}
    c=\langle \hat X_A \hat X_B \rangle ~.
   \label{eq_def_c_2qubits}
\end{equation}
Can we explain these data with a separable state, or is entanglement in fact needed to reproduce them? This problem can be formulated geometrically: if one plots the data ${\cal D}=\{m,c\}$ as a point on the plane, does it fall inside the region of points attainable by separable states? Or does it fall outside, in region of points only attainable by entangled quantum states? An important property, encountered throughout this review, is that these regions of points are \textit{convex} sets. That is: if ${\cal D}$ and ${\cal D}'=(m',c')$ are both attainable by separable states, then the any point $u{\cal D} + (1-u){\cal D}'$ (with $0\le u \le 1$) is also attainable by separable states. Indeed, if we denote as $p$ and $p'$ the corresponding probability distributions as in Eq.~\eqref{eq_rho_AB_sep_intro}, then $up + (1-u)p'$ is also a valid probability distribution describing a separable state; and gives rise to the data set $u{\cal D} + (1-u){\cal D}'$. Likewise, the region accessible to quantum states forms a convex region. Clearly, these regions are also bounded, since $|m|,|c|\le 1$. Nonetheless, as $\hat Z_A + \hat Z_B$ and $\hat X_A \hat X_B$ do not commute, not all values of $\{m,c\}$ can be achieved: for instance, if $m=1$ then the quantum state must be $|\uparrow\rangle\otimes|\uparrow\rangle$, for which $c=0$.\\

\noindent\textit{Bounding correlations as a ground-state problem.--}
In order to find the regions in the $(m,c)$ plane accessible to separable or general quantum states, it is therefore necessary and sufficient to find, for all directions $\vec w = (h, J)$ on the plane, the maximum and minimum of
\begin{equation}
    \vec w \cdot {\cal D} = \left\langle \frac{h}{2} (\hat Z_A + \hat Z_B) + J \hat X_A \hat X_B  \right\rangle ~.
    \label{eq_EW_as_GS_problem}
\end{equation}
In this expression, $\hat H(h,J) = \frac{h}{2} (\hat Z_A + \hat Z_B) + J \hat X_A \hat X_B$ can be formally interpreted as a quantum Hamiltonian (in this specific case, an Ising model in a transverse field for two qubits). If the mean value of $\hat H$ is taken over the set of separable states (resp. quantum states), by varying the parameters $(h, J)$ one reconstructs the region these states can span. Because of convexity, the boundaries of these regions are always achieved by \textit{pure} states. Furthermore, one can take into account the fact that separable pure states are simply \textit{product states}, namely states of the form $|\vec a\rangle\otimes|\vec b\rangle$ with $\vec a$ and $\vec b$ unit vectors on the Bloch sphere. The problem is therefore equivalent to finding the ground state of $\pm\hat H(h,J)$, either restricting to product states (that is, in a so-called ``mean-field'' approximation in the language of statistical physics), or over all quantum states (by diagonalizing $\hat H$). In the case of product states, we have $\langle \hat H(h,J) \rangle = (h/2)(z_A + z_B) + Jx_A x_B$, with $z_i, x_i$ the coordinates of the Bloch sphere vectors $\vec a$ and $\vec b$. Solving this minimization problem is an elementary exercise, and we find the complete set of constraints
\begin{align}
    m^2 + |c| \le 1  \qquad \text{for separable states,} \label{eq_wit_2qubits} \\
    m^2 + c^2 \le 1 \qquad \text{for quantum states.}
\end{align}
This is illustrated geometrically in Fig.~\ref{fig:2qubits_sep}. As it should be, the region accessible to quantum states is strictly larger than the region accessible to separable states. Equation~\eqref{eq_wit_2qubits} is called an \textit{entanglement witness}, namely an inequality that is valid for all separable states, and whose violation by some experimental data constitutes a proof that entanglement is present in the state under investigation \footnote{
In this specific case, Eq.~\eqref{eq_wit_2qubits} turns out to equivalent to the well-known Wineland spin squeezing condition \cite{wineland1992} $\langle \hat J_z \rangle^2 \ge N\langle \hat J_x^2 \rangle$, with $N=2$ and $\hat J_z = (1/2)\sum_{i=1}^N \hat Z_i$ is the collective spin along $z$ (and likewise for $\hat J_x$).
}. In generic many-body scenarios, one has often access to more than two mean values, and furthermore one cannot solve exactly the corresponding ground state problems. Nonetheless, the geometric structure of the problem, as well as the concept of entanglement witness, remain applicable, forming the basis of entanglement detection and characterization in many-body systems as well.\\

\begin{figure}
    \centering
    \includegraphics[width=1.\linewidth]{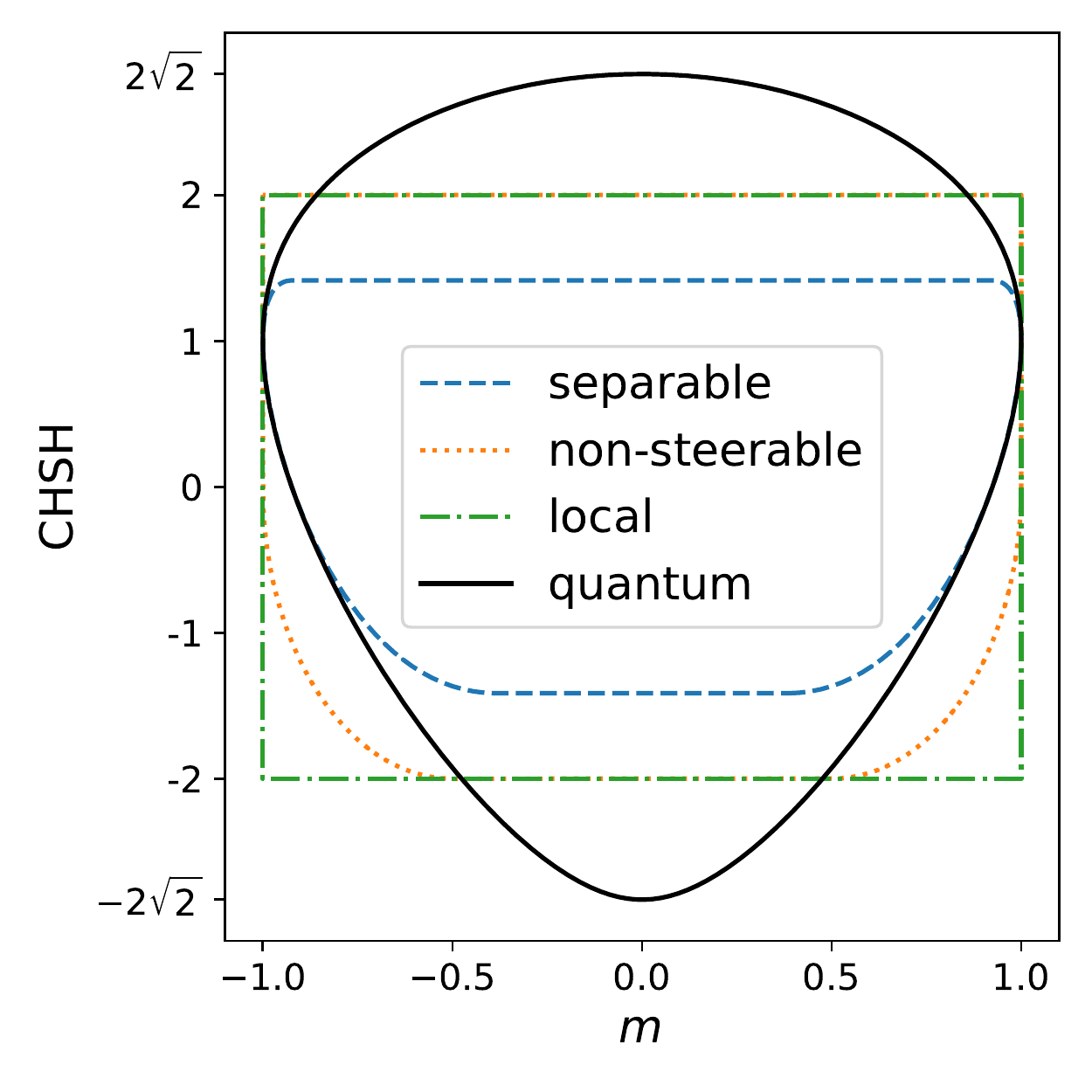}
    \caption{\textbf{Bell nonlocal correlations and EPR paradox versus versus separable and entangled quantum states.} Considering measurements of $m$ and CHSH as defined in Eqs.~(\ref{eq_m_nloc},\ref{eq_chsh_nloc}), the green dashed-dotted rectangular region delimits the data accessible to Bell local models; the orange dotted line to data accessible to local hidden state models (\ie non-steerable); the blue dashed line (resp.~black solid line) delimits the region accessible to separable quantum states (resp.~general entangled states) of two qubits by measuring the qubit observables $\hat X$ and $\hat Z$ for both Alice and Bob.}
    \label{fig:2qubits_nloc}
\end{figure}

\noindent\textit{Bell nonlocality.--} It is known that quantum mechanics allows for a form of correlations stronger than entanglement, namely Bell nonlocality \cite{brunneretal_RMP_2014}. While the former is revealed through an entanglement witness (or a separability criterion), nonlocality is revealed by the violation of a Bell inequality. As we will see, these inequalities can be seen as a special case of entanglement witnesses where \textit{no assumptions are made on the observables that are being measured}. This point of view is especially relevant in the context of this review, since as argued in the introduction a \textit{device-independent} characterisation of the system's quantum properties is desirable.

As a simple example, let us consider a bipartite scenario where at every repetition of the experiment Alice (resp. Bob) decides to perform either measurement $X^{A(B)}$ or $Z^{A(B)}$ on their system, and obtains as a result either $+1$ or $-1$. In practice, these measurements could be Pauli measurements $\hat X_{A(B)}, \hat Z_{A(B)}$ performed on a spin-$1/2$, and thus giving as a result spin up/down, but it is important to emphasize that the following discussion does not require this association.
According to Bell's locality assumption, where each system admits a description in terms of microstates, we may simply treat the above measurement outcomes as classical variables $(x_A, z_A, x_B, z_B) \in \{-1, +1\}^4$, corresponding to physical properties which are revealed by the act of measurement. In particular, according to this locality assumption, one is able to simultaneously assign values $\{x_A, z_A\}$ to the outcomes of possibly incompatible quantum observables, something which is forbidden by quantum mechanics. 

For the sake of illustration, let us consider that Alice and Bob use their results to compute the quantities
\begin{equation}
    m = (\langle z_A \rangle + \langle z_B \rangle) / 2  \label{eq_m_nloc}
\end{equation}
and
\begin{equation}
    {\rm CHSH} = \langle x_A x_B \rangle + \langle x_A z_B \rangle + \langle z_A x_B \rangle - \langle z_A z_B \rangle ~.
    \label{eq_chsh_nloc}
\end{equation}
Here, ${\rm CHSH}$ is the quantity entering the celebrated Bell inequality $|{\rm CHSH}| \le 2$, first discovered by Clauser, Horne, Shimony and Holt \cite{CHSH}. 
The question we want to answer is then: can these data be explained by a local model?

Geometricaly, the region of points $\{m,{\rm CHSH}\}$ accessible to Bell's local models forms a convex set, whose extremal points are given by the $2^4$ possible combinations of the local classical variables $\pm 1$. Therefore, this region is more precisely a polytope, called the \textit{local polytope}. 
The local polytope for our bipartite example is illustrated in Fig.~\ref{fig:2qubits_nloc}, together with the regions achievable with separable state or quantum states. For these latter two families, it is assumed that the measurements do correspond to performing Pauli measurements $\hat X$ and $\hat Z$ on a spin-$1/2$; and we solved numerically the minimization problems analogous to Eq.~\eqref{eq_EW_as_GS_problem}. The faces of the local polytope correspond to (tight) \textit{Bell inequalities}. In particular, the CHSH inequality $|{\rm CHSH}| \le 2$ turns out to be the only non-trivial inequality in the scenario we considered here (the inequality $|m|\le 1$ being trivial). As Bell's local models make no assumptions on the observables, they lead to a larger set of possible values for $\{m,{\rm CHSH}\}$ as compared to separable states. Therefore, the region accessible to separable states is \textit{strictly contained} inside the region accessible to Bell's local model, as illustrated in Fig.~\ref{fig:2qubits_nloc}. This property is completely general, and is what makes Bell inequalities (\ie device-independent entanglement witnesses) harder to violate than (device-dependent) entanglement witnesses. On the other hand, there are data points achievable by quantum states that cannot be reproduced by a Bell's local model, an example being the value ${\rm CHSH}=2\sqrt{2}$ obtainable by measurements on a maximally entangled pair $(|\uparrow \uparrow \rangle + |\downarrow \downarrow \rangle)/\sqrt{2}$ after local rotations of the qubits. Nevertheless, let us emphasize that Bell nonlocality is a form of quantum correlations strictly stronger than entanglement, as there are entangled states that admit a local model \cite{acinetal2006}.\\

\noindent\textit{Einstein-Podolsky-Rosen steering.--}
The concept of EPR steering \cite{wisemanetal2007} captures a manifestation of quantum correlations which is intermediate between entanglement and nonlocality. Following Einstein, Podolsky and Rosen seminal insight \cite{EPR}, EPR steering is fundamentally associated with the apparent violation of the Heisenberg uncertainty relations on Bob's system, when his measurements are conditioned on the outcomes of other measurements performed on Alice's system, however remote the two systems are. In the context of this pedagogical introduction, one may simply define EPR steering as the incompatibility of the data with a hybrid model, where $A$ is treated in a device-independent way (as in Bell's local models), while $B$ is treated as a quantum system with a well-defined state $|\vec b\rangle$ (as in separable states). Keeping the same example as for Bell nonlocality, in Eqs.~\eqref{eq_m_nloc} and \eqref{eq_chsh_nloc} $x_A,z_A$ are now $\pm 1$ variables, while $x_B, z_B$ are the components of a unit vector on the Bloch sphere. As illustrated in Fig.~\ref{fig:2qubits_nloc}, the region accessible to such hybrid models is strictly larger than the region accessible to separable states, and strictly smaller than the region accessible to Bell's local models. This reflects the hierarchy existing between these three types of correlations, which will be discussed further below.\\

In some sense, both EPR steering and Bell nonlocality may be viewed as stronger manifestations of quantum entanglement. All these three concepts aim at capturing the incompatibility of quantum correlations with a classical notion of microstate, according to which every subsystem has a ``well-defined'' state of some sort. They differ in the assumptions made about the subsystems, which result in a hierarchy among them. As illustrated in Fig.~\ref{fig:hierarchy}, entanglement is necessary but insufficient to have EPR steering, and EPR steering is necessary but insufficient to have Bell nonlocality. Importantly, this hierarchy holds not only in a bipartite case, but also in multipartite scenarios. Because of this structure, these three forms of quantum correlations result in inequivalent resources for quantum information tasks. Characterizing them in a given multipartite quantum system sheds lights on related but different non-classical properties of the same device.\\

\noindent\textit{Quantum correlations between two qubits.--}
One might think that in the simplest non-trivial case of two spin-$1/2$ particles, or qubits, everything is known on the separability problem. As a matter of fact, if the full density matrix $\hat\rho_{AB}$ is given, necessary and sufficient entanglement criteria are well-known, for instance based on the concurrence or on the positivity of the partial transpose (see \cite{horodecki_RMP_2009,GuhneT_2009} for reviews). Yet, several basic open questions remain, especially regarding the link between quantum entanglement and Bell's nonlocality, or the advantage in using generalized measurements (POVMs) over projective  measurements \cite{brunneretal_RMP_2014}. To give a particularly simple example, consider the so-called Werner state, namely the statistical mixture of a Bell pair with white noise: $\hat \rho_\eta = \eta |\Psi_-\rangle\langle\Psi_-| + (1-\eta)\mathbb{1}/4$, with $|\Psi_-\rangle=(|\uparrow\downarrow\rangle - |\downarrow\uparrow\rangle)/\sqrt2$. The Werner state is entangled if and only if $\eta > 1/3$, and if one uses projective measurements only, no Bell's inequality can be violated for $\eta\lesssim 0.71$ \cite{acinetal2006}; yet, it is not known if using generalized measurements, a Bell's inequality can be violated in this regime. This illustrate the complex relationship between the different notions of non-classical correlations already in the simplest case of two qubits. Nonetheless, despite the lack of a complete understanding, several general approaches to detect and probe entanglement from the partial knowledge of the quantum state of a many-body system are presented in this review. \\

\noindent\textit{Entanglement quantification.--}
Given that quantum entanglement plays a key role in many applications, one is naturally interested in \textit{quantifying} the entanglement content of a given system \cite{horodecki_RMP_2009}. For the simple case of two qubits, there is a unique way to order quantum states according to their entanglement content. A popular measure is given for instance by the so-called \textit{concurrence} (see ref.~\cite{horodecki_RMP_2009} for a review). The uniqueness of entanglement quantification in the two-qubits case can be traced back to the existence of a unique maximally-entangled state (namely, up to local unitaries, a Bell pair or spin singlet). However, especially in a many-body scenario where a unique maximally-entangled state does not exist, many inequivalent ways to formalize the quantification of entanglement emerge. Basically, for every task in which some form of quantum entanglement is a resource, one may quantify entanglement according to the performance in achieving the said task. Because of this, if a state $\hat \rho_1$ turns out to be more powerful than $\hat \rho_2$ in achieving a task $T$, according to the corresponding entanglement measure $E(\hat \rho_1) > E(\hat \rho_2)$, while if at the same time $\hat \rho_2$ is more powerful than $\hat \rho_1$ for another task $T'$, according to the corresponding entanglement measure $E'(\hat \rho_2)>E'(\hat \rho_1)$. From this perspective, it should not be a surprise that no entanglement measure is fundamentally better than all other ones. Yet, it is commonly agreed that any entanglement measure $E$ must fulfill a minimal list of formal properties, such as
\begin{enumerate}
    \item Positivity: For all quantum states $\hat \rho$, $E(\hat \rho) \ge 0$ with $E(\hat \rho_{\rm sep})=0$ for all separable states $\hat \rho_{\rm sep}$.
    \item Monotonicity under local operations and classical communications (LOCC): if $\hat \rho_2$ is obtained from $\hat \rho_1$ by acting locally on the subsystems, possibly assisted by classical communication to coordinate the local actions, then $E(\hat \rho_2) \le E(\hat \rho_1)$. 
\end{enumerate}
These reflect the idea that entanglement should be defined as a resource which is not present in separable states (property 1), and which cannot be created using only classical means of communication (property 2). Naturally, one may also aim at quantifying EPR steering and Bell's nonlocality in analogous frameworks, although this goes beyond the scope of this pedagogical section.

\subsection{Multipartite quantum correlations}
\subsubsection{Introduction to the problem}
Consider a system composed of $N \gg 1$ components, or subsystems. Concretely, these can be individual neutral atoms, trapped ions, optical modes, superconducting qubits, NV centers in diamond, localized spins in condensed-matter systems, etc. Or the subsystems could also be larger groups of these individual components, such as a collection of $N$ atomic ensembles each composed by many particles. Our goal is to reveal and characterize the quantum nature of the correlations within the system, without having to perform full-state tomography. In fact, when $N\gg 1$ it becomes experimentally unfeasible to reconstruct the full quantum state $\hat \rho$ of the global system, either because the required measurements cannot be implemented in practice, or simply because the number of required measurements, which scales exponentially with $N$, is too large. In practice, thus, we only have partial information about $\hat \rho$. As an example, let us imagine that some local observables $\hat X_i, \hat Y_i, \dots$ can be measured, where the label $i\in\{1,\dots,N\}$ denotes the subsystems (for instance, if the subsystems are qubits, $\hat X_i, \hat Y_i, \dots$ could be spin measurements along different directions; notice that in order to probe entanglement, it is necessary that different non-commuting observables can be measured on each subsystem). Then, assume that some statistical correlations between different subsystems are observed, e.g. 
\begin{eqnarray}
    \langle \hat X_i \hat X_j \rangle &\neq& \langle \hat X_i \rangle \langle \hat X_j \rangle ~, \\
    \langle \hat Y_i \hat Y_j \rangle &\neq& \langle \hat Y_i \rangle \langle \hat Y_j \rangle ~,~{\rm etc.}
    \label{eq_correlations}
\end{eqnarray} 
for some pairs of subsystems $i \neq j$. Here, $\langle \hat X \rangle = {\rm Tr}(\hat \rho \hat X)$ is the quantum expectation value in the unknown state $\hat \rho$. What can we say about the presence of quantum correlations just from these very incomplete observations? Are quantum correlations of any kind at all required in order to explain the observed statistics? Or can we instead invoke a purely ``classical'' explanation for the correlations observed? Moreover, if we are able to conclude that the measured data require the subsystems to share quantum correlations, then what further insight can be gained about the system's state; can we extract more qualitative and quantitative information about the structure of these correlations? Or, can they be used as a resource for some quantum-technology application? As we shall see, formalizing these questions into well-defined mathematical problems lead to complementary perspectives on detecting and characterizing quantum correlations in many-body systems. \\

\subsubsection{Quantum entanglement}
A multipartite quantum state is said to be \textit{entangled} if it cannot be decomposed as 
\begin{equation}
    \hat \rho_{\rm sep} = \sum_\lambda p_\lambda \otimes_{i=1}^N \hat \rho_i^{(\lambda)} \;,
    \label{eq_rhosep}
\end{equation}
where $p_\lambda \ge 0$ is a probability distribution ($\sum_\lambda p_\lambda=1$), and $\hat \rho_i^{(\lambda)}$ is an arbitrary quantum state for subsystem $i$. A state of the form $\hat \rho_{\rm sep}$ is called \textit{separable}, and such states form a convex set. In a separable state, all subsystems have a well-defined local quantum state (LQS) $\hat \rho_i^{(\lambda)}$; and these LQS are jointly sampled with probability $p_\lambda$. As the LQS of different subsystems might not be independent from each other, some correlations may anyways be present in a separable state. These have, however, a classical explanation as compared to the correlations allowed by more general (entangled) quantum states. 

When performing an experiment, a measurement labeled by $x_i$ is performed on particle $i$, giving result $a_i$. Correlations in the multipartite system are then investigated from the joint probability distribution $p[\Vec{a}\vert\Vec{x}]$, describing the probability of obtaining results $\Vec{a}=\{a_1,a_2, ...,a_N\}$ when measurements $\Vec{x}=\{x_1,x_2, ...,x_N\}$ are performed. For a separable state this probability distribution can be written as
\begin{equation}
    p_{\rm sep}[\vec a|\vec x] = \sum_\lambda p_\lambda \prod_{i=1}^N p_i^Q(a_i|x_i,\lambda)  \;,
    \label{eq_SeparableModel}
\end{equation}
with local probabilities admitting a quantum description (hence the superscript $Q$) through the Born rule $p_i^Q(a|x,\lambda) = {\rm Tr}[\hat \rho_i^{(\lambda)} \hat E_i^{(x,a)}]$, where $\hat E_i^{(x,a)}$ is the POVM associated to obtaining result $a$ when measurement $x$ is performed on system $i$ (remember $\hat E_i^{(x,a)} \ge 0$, and $\sum_a \hat E_i^{(x,a)} = 1$). Therefore, if a set of experimental data cannot be explained by a probability distribution of the form Eq.~\eqref{eq_SeparableModel}, we have to conclude that the underlying quantum state is entangled.\\

\begin{figure}
    \centering
    \includegraphics[width=.8\linewidth]{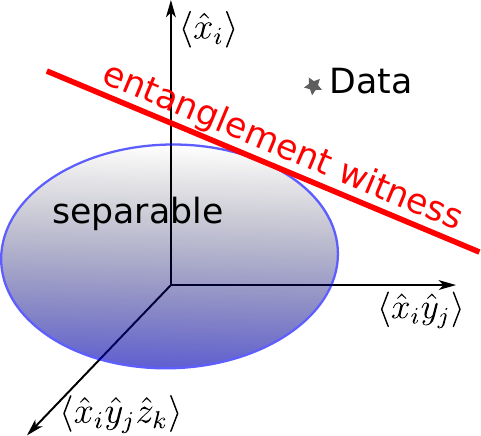}
    \caption{\textbf{Entanglement witness.} Data obtained from measurements on separable states form a convex set in the space of the measured correlators (shaded region). On the other hand, measurements on entangled states can give rise to data outside of this region (star). In this geometrical picture, entanglement witnesses are hyperplanes defining half-spaces that contain the set of separable data. For this reason, they allow for determining whether some observed data require entanglement. In the illustration, $\hat x_i, \hat y_j, \hat z_k$ are some observables measured on subsystems $i,j,k$.}
    \label{fig_schema_EW}
\end{figure}

\noindent\textit{Entanglement witnesses.--} How can we prove that no separable state can reproduce some given experimental data? As an example, let us assume that these data are average values of some observables $\hat O_1, \dots, \hat O_n$ (\eg $\hat O_1 = \hat X_1 \hat X_2$, $\hat O_2 = \hat X_1 \hat Y_3$, $\hat O_3=\hat Z_2$, etc.) Then, this data form the set ${\cal D}=\{\langle \hat O_a \rangle \}_{a=1}^n$, which geometrically is a point in a $n$-dimensional space. By definition, we say that ${\cal D}$ is a \textit{separable data set} if it can be reproduced by a separable state, Eq.~\eqref{eq_rhosep}. Entanglement detection thus consists in answering the question: how can we prove that ${\cal D}$ is not a separable data set? A crucial property is that separable data sets form a convex set. This property is elementary, and follows simply from the convexity of the set of separable quantum states, Eq.~\eqref{eq_rhosep}. In order to prove that ${\cal D}$ is \textit{not} a separable data set, it is sufficient to find a hyperplane separating ${\cal D}$ from the set of separable data sets. Such a separating hyperplane is then called an \textit{entanglement witness}, suitable to detect entanglement from the data set ${\cal D}$, see Fig.~\ref{fig_schema_EW}. Geometrically, a hyperplane is defined by its normal vector $\vec w = (w_1, \cdots, w_n)$, such that the corresponding entanglement witness takes the form
\begin{equation}
    \langle \hat W \rangle := \sum_{a=1}^n w_a \langle \hat O_a \rangle - B_{\rm sep} < 0 ~,
    \label{eq_EW}
\end{equation}
where $B_{\rm sep}$ is the \textit{separable bound} of $\sum_{a=1}^n w_a \hat O_a$, that is, the minimal value of $ \sum_{a=1}^n w_a {\rm Tr}(\hat O_a \hat \rho_{\rm sep})$ over all separable states, Eq.~\eqref{eq_rhosep}. Notice that, by virtue of the convexity of the set of separable states, this minimum is always achieved by a pure state, namely a product state of the form $|\Psi\rangle = \otimes_{i=1}^N |\psi_i\rangle$. For any non-separable data set there always exist witnesses allowing one to conclude that entanglement is present. On the contrary, let us emphasize that entanglement witnesses are only sufficient conditions, since a given witness might correspond to an hyperplane that fails in separating the data under investigation from the set of separable data. For this reason, a major difficulty in many-body scenarios is to find entanglement witnesses that are suitable to detect entanglement for given states and measurements. Some techniques to tackle this problem will be presented in this review. \\

\noindent\textit{Multipartite entanglement structure.--} If an entanglement witness is violated, this implies that the quantum state under consideration cannot be decomposed as Eq.~\eqref{eq_rhosep}. In general, however, this does not provides much information about the structure of entanglement within the system. For example, it could be the case that subsystems 1 and 2 are entangled but both uncorrelated from the other subsystems, the latter being described by a fully separable model. In order to provide further insight into the entanglement structure, one may consider different partitionings of the subsystems $\{1, \dots ,N \}$ into subgroups ${\cal G} = \{G_1, \dots, G_Q\}$, where $G_q$ are non-overlapping groups comprising $|G_q|$ subsystems, such that $\cup_{q=1}^Q G_q = \{1, \dots, N\}$. Equation~\eqref{eq_rhosep} thus corresponds to the special partitioning into individual subsystems ($|G_q|=1$ for all $q$). 

We say that a state is separable with respect to the partitioning ${\cal G}$, or in short ${\cal G}$-separable, if it can be decomposed as $\rho_{{\cal G}-{\rm sep}} = \sum_\lambda p_\lambda \otimes_{G \in {\cal G}} \hat \rho_G^{(\lambda)}$, where $\hat \rho_G^{(\lambda)}$ is a quantum state for the subsystems within the group $G$. The major difference with Eq.~\eqref{eq_rhosep} is that quantum correlations within each group are now possible.

We define the \textit{depth} of the partitioning ${\cal G}$ as the size of its larger subgroup, ${\rm depth}({\cal G}) = \max_q |G_q|$. We say that a quantum state is $k$-producible if it can be decomposed as a statistical mixture of ${\cal G}$-separable states whose depth is at most $k$. By definition, a state which is not $k$-producible has an entanglement depth of at least $k+1$. On the other hand, we say that a quantum state is $k$-separable if it can be decomposed as a statistical mixture of ${\cal G}$-separable states such that $|{\cal G}| \ge k$ (that is, the partitioning ${\cal G}$ is composed of at least $k$ groups of subsystems). The notions of $k$-producibility and $k$-separability are partly complementary (for instance, if a state is $1$-producible, it is also $N$-separable, also called fully-separable, corresponding to the situation of Eq.~\eqref{eq_rhosep}). More recently, alternative characterizations of multipartite entanglement involving combinations of $k$-producibility and $k$-separability have been investigated. Crucially, all these families of quantum states form convex set, that is, the statistical mixture of two states in a given family belongs to the same family. Consequently, the corresponding data sets also form convex sets (in the space of data sets), which can be conveniently characterized by suitable entanglement witnesses of the same form as Eq.~\eqref{eq_EW}, albeit with a bound which depends on the specific class of entangled states which one aims at certifying. We refer the reader to ref.~\cite{Szalay2019,renetal2021} for recent results on such notions regarding the structure of multipartite entanglement. \\

\noindent\textit{Entanglement quantification.--}
While $k$-producibility and $k$-separability provide a useful characterization of the entanglement structure of the system, they do not provide any direct information on the ``strength'' of these quantum correlations, for example in terms of robustness to noise or usefulness for a specific quantum information task. A correct quantification of the nonclassical resources present in a state requires the definition of appropriate entanglement measures. The latter are chosen to be nonnegative real functions $\mathcal{E}(\rho)$ such that~\cite{VedralPlenio98,PlenioVirmani07}: i) $\mathcal{E}(\sigma)=0$ for all separable states $\sigma$; and ii) $\mathcal{E}(\rho)$ does not increase on average under local operations and classical communication (LOCC) \footnote{We note that this differs from the convention in, e.g.,~\cite{PlenioVirmani07}, where such quantities are referred to as `monotones'.}. As we already emphasized, based only on these two requirements, many inequivalent measures can be defined, each resulting in a different ordering of the entangled states. For this reason, it is usually favorable to consider measures that are associated to questions of practical relevance, such that they inherit a concrete meaning.

In the bipartite case, typically adopted measures are entropies, Schmidt rank, concurrence or entanglement of formation/distillation \cite{PlenioVirmani07,friietal_review_2018}. In the multipartite case, however, even more possibilities arise, reflecting the complexity of multipartite LOCC classification and the lack of a unique maximally entangled state \cite{DeVicenteSpeeKraus2013,SauerweinWallachGourKraus2018}. These measures are often operationally related to communication tasks, where entanglement is seen as a resource for transcending the limitations of LOCC for distant parties. For many-body systems, however, other features are typically more relevant. One example is the best separable approximation (BSA)~\cite{LewSanp98,KarnasLew01}, which quantifies how closely a many-body state can be approximated by a separable state. Another example is the generalized robustness (GR)~\cite{SteinerPRA03}, which quantifies the robustness to noise of a state, in terms of the minimal amount of (arbitrary) noise to be added in order to make the state separable. 

Unfortunately, the exact computation of entanglement measures is often unfeasible, as they require full state tomography, which is in practice inaccessible for many-body states. For this reason, strategies have been developed to at least bound entanglement measures of interest from below, based on a partial knowledge of the system state. For example, it has been shown that there are methods to relate the violation of entanglement witnesses to lower bounds for the BSA and GR \cite{FadelPRL21}.

We refer the reader to ref.~\cite{friietal_review_2018} for an in-depth review on entanglement quantification and its practical implementation in experiments.

\subsubsection{Bell nonlocality} 
While the presence of entanglement rules out a description of the observed statistics in terms of local probability distributions admitting a quantum description, see Eq.~\eqref{eq_SeparableModel}, Bell nonlocality rules out a description in terms of local probability distributions of \textit{any} kind.
For this reason, Bell nonlocality is a stronger form of correlations than entanglement, and actually the strongest known form of correlation observed to appear in nature \cite{brunneretal_RMP_2014}.

Formally, probability distributions that admit a local description can be written as
\begin{equation}
    p_{\rm local}[\vec a|\vec x] = \sum_\lambda p(\lambda) \prod_{i=1}^N p_i(a_i|x_i,\lambda)   \;,
    \label{eq_BellLocalModel}
\end{equation}
where, contrary to Eq.~\eqref{eq_SeparableModel}, $p_i(a_i|x_i,\lambda)$ are arbitrary probability distributions. By definition, if a set of experimental data
cannot be explained by a probability distribution of the
form Eq.~\eqref{eq_BellLocalModel}, we say that it demonstrates Bell nonlocality.

In the context of Eq.~\eqref{eq_BellLocalModel}, $\lambda$ is called a \textit{local hidden variable}, and it allows one to take into account that correlations might originate from unknown common causes or unobserved degrees of freedom. Crucially, the local probabilities $p_i(a_i|x_i,\lambda)$ might depend on $\lambda$, but they do not depend on the measurement setting or result for another system $j\neq i$. Even though the local probabilities could be \textit{a priori} random functions of $\lambda$, it turns out that all randomness can be formally incorporated in $p(\lambda)$, so that to find a local model it is sufficient to consider only \textit{deterministic} functions $p_i(a_i|x_i,\lambda)$. Namely, one can simply consider that in each repetition of the experiment, every subsystem has a pre-determined outcome $a_i$ for each possible settings $x_i$; and that this ``pre-existing'' outcome is simply ``revealed'' by the measurement. In a Bell local model, one can then simply attach local classical variables $a_i^{(x)}$ to represent the outcomes on each subsystem, and assume that such local variables are randomly sampled at each repetition of the experiment.

To see that Bell nonlocality is a form of quantum correlations stronger than entanglement, it is easy to verify that separable quantum states [Eq.~\eqref{eq_rhosep}] always admit a local model [Eq.~\eqref{eq_BellLocalModel}], since one can consider $p_i(a|x,\lambda) = {\rm Tr}[\hat \rho_i^{(\lambda)} \hat E_i^{(x,a)}]$. This means that, if a state is separable, then no Bell inequality can be violated; and this holds \textit{regardless of any assumption about the Hilbert space dimension or the actual POVM which were implemented}. Such crucial property implies that Bell inequalities can be used as \textit{device-independent} entanglement witnesses: they allow one to detect quantum entanglement while relaxing assumptions about the precise calibration of the measurements, and about the actual Hilbert space explored by the subsystems. In fact, let us emphasize that in the context of a Bell test, there is no need to give a description for the state of the system or the measurement performed (\eg through quantum states and operators). Bell inequalities are based on probability distributions of obtaining results $a_i, a_j, \dots$ when measurements labeled as $x_i, x_j, \dots$ are performed on subsystems $i, j, \dots$, without making any assumption on what these measurements actually are. It is largely a choice of the experimenter to assign such labels, for example they can label a Pauli $\hat Y$ measurement $x=1$ and the spin up/down results $a=\pm 1$. Moreover, each party can perform completely different measurements (such as spin, position, etc.) and apply different choices of coarse-graining and labeling of the results to organize their results.

\noindent\textit{Multipartite Bell inequalities.--} How can one prove that no local model [Eq.~\eqref{eq_BellLocalModel}] can reproduce some experimental data? Formally, this problem is very similar to the one of finding a violated entanglement witness. In a realistic many-body scenario, one cannot reconstruct the full joint probability distribution $p[\Vec{a}|\Vec{x}]$, as this would require implementing exponentially-many different experimental settings, each one repeating many times in order to gather enough statistics to estimate the corresponding probabilities. In practice, one rather probes correlations among subsystems through coarse-grained data, such as \eg marginals of the full $p[\Vec{a}|\Vec{x}]$, correlations among only pairs of subsystems, or more generally through some function of the input/output settings, averaged over many repetitions of the experiment. Using the variables $a_i^{(x_i)}$ to denote the outcome on the $i$th subsystem when the measurement setting $x_i$ is implemented, such averaged functions can be written as $\langle f_i[\{a_i^{(x_i)}\}]\rangle$. A set of experimental data is thus ${\cal D} = \{\langle f_1[\{a_i^{(x_i)}\}] \rangle, \dots, \langle f_n[\{a_i^{(x_i)}\}] \rangle\}$, which geometrically is also a point in a $n$-dimensional space. Similarly to the case of entanglement, we define \textit{local data sets} as data sets which can be reproduced by a local model Eq.~\eqref{eq_BellLocalModel}, such that the problem of nonlocality detection can now be formulated as the task of proving that a collection of measurements is not a local data set. To this goal, note that the set of local data sets is convex, and even more that it is a polytope.
\textit{Bell's inequalities} are then defined geometrically as hyperplanes separating local data sets from nonlocal ones, and in the case where they coincides with the facets of the local polytope they are said to be \textit{tight}. A tight Bell's inequality can be written as:
\begin{equation}
    \langle {\cal B} \rangle := \sum_{k=1}^n w_k \langle f_k[\{a_i^{(x_i)}\}] \rangle - B_{\rm loc} < 0 ~,
    \label{eq_BI_generic}
\end{equation}
where $B_{\rm loc}$ is the \textit{local bound} of $\sum_{k=1}^n w_k f_k[\{a_i^{(x)}\}]$, that is, the minimal value of $\langle {\cal B} \rangle$ over all possible values of the local variables $\{a_i^{(x_i)}\}$. The main challenge in a many-body scenario is to identify Bell inequalities that are suitable to detect nonlocality for a given states and measurements.\\

\textit{Further considerations on multipartite nonlocality.--}
Equation \eqref{eq_BellLocalModel} describes a fully-local model, in which every subsystem is associated to a probability distribution. This is the analog, in the context of Bell nonlocality, to what is a fully-separable state in the context of entanglement, \ie Eq.~\eqref{eq_rhosep}. Similarly to the case of entanglement detection, violating a multipartite Bell inequality associated to such a fully-local model does not, in general, provide much information about the detailed structure of the nonlocal correlations within the multipartite system. It might be the case, for instance, that subsystems 1 and 2 form a Bell pair fully uncorrelated from the other subsystems, the latter being described by a fully local model. Several works in the recent years have therefore proposed to derive Bell's inequalities detecting the finer structure of nonlocality (see ref.~\cite{brunneretal_RMP_2014} for a review). In essence, such works discuss notions which parallel the concepts of $k$-producibility and $k$-separability presented in the previous section, with the local quantum states being replaced by local probability distributions. Given a partitioning ${\cal G}=(G_1, \cdots, G_Q)$ of the $N$ subsystems into $Q$ partitions, as defined in the previous section, one defines a ${\cal G}$-local model as $p_{{\cal G}-{\rm local}}[\vec a | \vec x] = \sum_\lambda p(\lambda) \prod_{G \in {\cal G}} p_G(\vec a_G | \vec x_G, \lambda)$, where $\vec a_G = \{a_i\}_{i \in G}$ are the outcomes for parties inside the $G$ partition, and $\vec x_G$ are similarly the corresponding inputs. Crucially, the probability distribution $p_G(\vec a_G | \vec x_G, \lambda)$ may describe nonlocal correlations \textit{within} the group $G$. One then considers statistical mixtures of such models for different partitioning, and introduce notions such as $k$-producibility and $k$-separability in full analogy with the case of entanglement \cite{bancaletal2009,courchodetal2015}. A probability distribution $p[\vec a |\vec x]$ is said to be genuine multipartite nonlocal if it cannot be decomposed in such a way, unless one includes all $N$ subsystems inside the same `partition'. Seminal studies on genuine multipartite nonlocality, and crucially, the possibility to detect it through suitable generalizations of Bell's inequalities, were done by Svetlichny for $N=3$ parties \cite{Svetlichny1987}, and subsequently generalized to arbitrary $N$~\cite{SeevinckPRL02,CGPRV02}. Later studies have suggested to impose further constraints on the intra-group probability distributions $p_G(\vec a_G | \vec x_G, \lambda)$, such as no-signalling constraints, to obtain more physical models (see ref.~\cite{brunneretal_RMP_2014} and references therein). In general, it was progressively recognized that the violation of a multipartite Bell inequality is able to provide a device-independent characterization of the underlying entangled state from the measurement statistics alone \cite{bancaletal2011,brunneretal2012}. More recently, this was also extended to an assumption-free characterization not only of the state, but also of the measurements performed, namely the so-called self-testing \cite{Supic2020selftestingof}. In the Sections on theoretical and experimental achievements, we will review recent results on probing the depth of Bell nonlocality, and the device-independent characterization of multipartite entanglement, through the violation of suitable Bell inequalities with many-body systems.

\subsubsection{Einstein-Podolsky-Rosen steering}
The notion of EPR steering, formally defined in ref.~\cite{wisemanetal2007}, is intended to capture the original insight of Einstein, Podolsky and Rosen \cite{EPR} that quantum mechanics seemingly allows one to instantaneously modify (``steer'') the state of a system by performing measurements on another system, irrespective of their spatial separation. For an extensive review on the topic, we refer the reader to refs.~\cite{reidetal_RMP_2009,uola_etal_RMP_2020}. In the context of this review, where our main focus is to present various notions of non-classical correlations and how they can be probed in many-body systems, EPR steering corresponds to a situation intermediate between quantum entanglement and Bell's nonlocality. In this situation, the ``classical" model is a hybrid one, where some subsystems are described by a local quantum state (these are the ``trusted" subsystems, $i \in {\cal T}$), and some others by local variables as in Bell's local models (these are the ``un-trusted" subsystems, $i \in \bar{\cal T}$). We can write this so-called local hidden state (LHS) model as \cite{cavalcantietal2011}:
\begin{equation}
    p[\vec a | \vec x] = \sum_{\lambda}p_\lambda \prod_{i \in {\cal T}} {\rm Tr}[\hat \rho_i^{(\lambda)} \hat E_i^{(x_i,a_i)}]~ \prod_{i \in \bar{\cal T}} p(a_i|x_i,\lambda) ~.
    \label{EPRlocalmodel}
\end{equation}
Therefore, exactly in the same way as for quantum entanglement and Bell nonlocality, also EPR steering, \ie the incompatibility of the data with a decomposition as in Eq.~\eqref{EPRlocalmodel}, can be probed through the violation of suitable inequalities. A prominent example is given by the detection of EPR steering through the violation of Heisenberg-like uncertainty relations \cite{ReidPRA89}, following the thought experiment by EPR. Alternatively, it is possible to derive steering inequalities numerically, from the solution of semidefinite programs aimed to rule out LHS models \cite{Cavalcanti_2017}.

At this point it is worth mentioning that, while the definition of EPR steering for a bipartite system is straightforward, its multipartite generalisation can become conceptually difficult. For example, due to the intrinsic asymmetry of these correlations, it is not obvious how to generalise the notion of genuine multipartite entanglement, or nonlocality, to steering. An approach could be to exclude LHS models for all (or some) possible permutations of which are the trusted/untrusted parties p\cite{heReid2013,tehReid2014,tehetal2022}. Another proposed approach consists in defining which are the trusted/untrusted parties, and then conclude genuine multipartite entanglement in this semi-device-independent scenario \cite{Cavalcanti_2015,Mattar_2017}. Moreover, in the multipartite case, so-called monogamy relations have been derived whose complete characterization is particularly challenging \cite{tehReid2014,Armstrong2015,tehetal2022}, and represents an important open problem in the field.

\section{Theoretical achievements}\label{sec3}
In this section, we review the main methods which have been developed in recent years to address the problem of characterizing quantum correlations in many-body systems. In particular, we will present so-called data-driven methods, which can be used to infer new criteria from the data themselves, as well as criteria involving low-order moments of collective observables which are by construction valid for arbitrary number of particles. We concentrate our attention mostly on methods whose complexity does not scale exponentially with the system size, and which can be applied to many-body systems of large size (beyond hundreds of qubits). We say that a method for probing quantum correlations is \textit{scalable} if the resources needed to its implementation scale at most polynomially with the number of particles considered. Here, by resources we mean e.g. the number of measurements needed in order to accumulate enough data in order to implement a given method, or to test a given criterion; or the time and memory required by a (classical) computer to run an algorithm deciding whether or not quantum correlations are needed to explain the obtained measurement results.

\subsection{Entanglement witnesses and Bell's inequalities from a given collection of mean values}

The central question: ``Can a given collection of measured mean values of observables on a multipartite system be reproduced without quantum correlations?'' has been addressed from various perspectives over the last decades. Seminal works have proposed a hierarchy of criteria involving partial transposition of the multipartite density matrix, or extensions thereof \cite{dohertyetal2002,dohertyetal2004,dohertyetal2005}. Although offering in principle a complete solution to the problem of entanglement detection, these approaches cannot be applied beyond a few qubits, as they involve numerical manipulations of the full density-matrix. Alternative approaches based on the knowledge of covariance matrices \cite{gittsovichetal2010}, or correlation tensors \cite{deVincenteH2011}, have also been proposed. Yet, it is not clear how such methods could be adapted when some entries of the covariance matrix or correlation tensors are unknown; furthermore, they offer only sufficient conditions for entanglement, so that the data could be incompatible with a separable state, yet entanglement is not detected by the said criteria. Systematic approaches tackling this problem in its full generality and in a scalable way were first developed in the context of Bell's inequalities. This was achieved by two complementary approaches, which both exploit the fact that this is intrinsically a problem of \textit{classical} statistical physics. 

The first approach \cite{baccarietal2017} is rooted in a mathematical framework called \textit{semi-definite programming}, and exploits the fact that the measured mean values (which we call ``data'' in this review for simplicity) can be written as entries in a covariance matrix (or linear combinations of the entries), which is semi-definite positive by construction. If one assumes that there exists an underlying Bell local model as in Eq.~\eqref{eq_BellLocalModel}, the entries of this covariance matrix obey extra constraints. Verifying that the data are compatible with such constraints can be solved with standard numerical routines of semi-definite programming; and when the data violate these conditions, one obtains an explicit Bell's inequality in the form of Eq.~\eqref{eq_BI_generic}, in which the coefficients $w_k$ have been inferred from the data themselves. The approach is intrinsically scalable, as one never deals with the complete Bell local model itself, which contains exponentially-many free parameters, but only with necessary conditions that the data must obey if this classical model exists. It is very flexible, as it can be adapted to correlation functions of arbitrary order without loss of scalability, as already illustrated in the seminal paper \cite{baccarietal2017}. An intrinsic limitation is that the method only considers necessary conditions for a Bell local model to exist: it can be the case that the data are compatible with these semi-definite positive constraints, but are in fact incompatible with a Bell local model; yet the method can be systematically tightened by considering covariance matrices of increasing order, forming a converging hierarchy of nonlocality detection tests -- but at a computational cost which grows rapidly. 

The second, complementary, approach \cite{frerotR2021_bell} is rooted in the framework of inverse statistical problems, and in particular so-called \textit{inverse Ising problems}. It is in essence a variational approach. The basic insight is that the most generic Bell local model of Eq.~\eqref{eq_BellLocalModel} can be parametrized, without loss of generality, as a Gibbs distribution of the form $
    p[\{a_i^{(x'_i)}\}] = Z^{-1}[{\bf K}]\exp\left(
        \sum_k K_k f_k[\{a_i^{(x'_i)}\}]
    \right)$.
In this expression, the functions $f_k$ represent the observables which compose the collection of mean values, and the $K_k$ are the variational parameters which are optimized in order to reproduce the observed data $\langle f_k \rangle_Q$. Three key features of this approach should be emphasized: first, it is \textit{complete} in that varying the parameters $K_k$ allows one to exactly map the data sets which can be achieved with the most general Bell local models. Second, optimizing the parameters $K_k$ is achieved using a \textit{convex} cost function $C({\bf K})$ (which is a generalized Gibbs free energy, endowed with suitable convexity properties), whose gradient is simply given by $\partial C/\partial K_k = \langle f_k \rangle_{\bf K} - \langle f_k \rangle_Q$,
where the first average is over the $p$ distribution, while the second average is the target data to reproduce. Notice that only the gradient of the cost function has to be computed in order to optimize the variational parameters $K_k$ in a gradient-descent algorithm, or any improvement thereof. The cost function $C$ itself need not be evaluated. The simple expression for the gradient leads to the third key property: as it involves only the statistical average of the $f_k$ functions over the $p$ distribution, it can be efficiently evaluated using standard tools of classical statistical physics, such as e.g. Monte Carlo methods, which are intrinsically scalable to many-body systems. When the data cannot be reproduced with a Bell local model, the optimization converges towards the closest approximation to them in the space of data; and from the distance between the target data and the optimal classical model, a violated Bell's inequality is reconstructed.

Interestingly, both approaches have been applied to many-body problems \cite{baccarietal2017,frerotR2021_bell}, and have delivered novel Bell's inequalities. While the coefficients of the inferred Bell's inequalities are obtained numerically, it was sometimes possible to recognize analytically certain patterns and symmetries in these coefficients, leading the authors to derive analytically the Bell's inequalities, their classical bound, sometimes even their quantum bound, and to generalize them to new families of Bell's inequalities tailored to many-body quantum states of interest \cite{baccarietal2017,baccarietal2020,frerotR2021_bell,frerotA2021}. 

In continuation of these seminal works to infer new Bell's inequalities form a given collection of mean values, similar approaches have been developed to infer new entanglement witnesses \cite{frerotetal2022,FrerotR_2021a} (see also \cite{Navascues2021entanglement}). In this context, the mathematical framework remains the same, but the underlying classical models correspond to separable quantum states [Eq.~\eqref{eq_rhosep}], where each local quantum state $\rho_i^{(\lambda)}$ is parametrized by a set of classical variables. With these approaches as well, the inferred coefficients of the entanglement witness [Eq.~\eqref{eq_EW}] are obtained numerically; yet it was often possible to analyse them analytically, yielding novel families of EW of general importance, beyond the specific data used to discover them \cite{frerotetal2022}. Furthermore, let us mention that once a specific entanglement witness has been inferred from the data, its classical bound might alternatively be found resolving a set of algebraic equations \cite{sperlingV2013,gerkeetal2018}.

These approaches could be extended in a straightforward manner to the hybrid situation where some subsystems are treated in a device-independent way, as for Bell's local models, and some others are described with a given Hilbert space and measurement operators. In this hybrid situation, which has never been explored with these techniques in a many-body context, one would derive EPR-steering inequalities tailored to a given set of many-body data.

The approaches of refs.~\cite{baccarietal2017,FrerotR_2021a,frerotR2021_bell,frerotetal2022} offer constructive, complete, and scalable solutions to the problem of entanglement detection from a collection of mean values measured on a many-body system. They offer a relevant paradigm to probe entanglement in generic, spatially structured, many-body systems from realistic data as obtained on present-day quantum simulators and computers. Yet their numerical implementation can be demanding, which motivates the development of complementary approaches, of less general applicability but also of less computational complexity.

\subsection{Entanglement witnesses and Bell inequalities from permutationally-invariant correlators}

Consider the situation where local observables $\hat O_a^{(i)}$ are defined on the $N$ subsystems. Here, $i=1,\dots, N$ labels the subsystem, while $a=1,\dots,K$ labels the different local observables. When the subsystems are qubits, the local observables are spin-1/2 observables, but can be arbitrary for generic qudits subsystems with local dimension $d>2$. We assume that experimentally, their mean value can be estimated by collective measurements, yielding the knowledge of $\sum_{i=1}^N \langle \hat O_a^{(i)} \rangle$. Similarly, we assume that two-body correlations of the form $\sum_{i\neq j} \langle \hat O_a^{(i)} \hat O_b^{(j)} \rangle$ can be inferred, at least for some of the pairs of measurement settings $(a,b)$. Notice that the data are invariant under any permutation of the subsystems. This framework includes, for instance, the measurement of first- and second-moment of collective spin observables $\hat J_a = \sum_i \hat S_a^{(i)}$ for spin-1/2 ensembles, as routinely performed in several cold-atoms experiment discussed in Section~\ref{sec4}. The basic question we aim at answering in this context is the following: can we prove from such data that entanglement is present among the subsystems? If only up to second order moments are known, this question was first completely solved in the context of spin-$1/2$ ensembles \cite{tothetal2009}, where a complete family of so-called generalized spin-squeezing inequalities was derived. These correspond to analytical conditions on the first ($\langle \hat J_a \rangle$) and second ($\langle \hat J_a \hat J_b \rangle$) moments of the collective spin along the three spatial directions ($a,b=x,y,z$), which detect entanglement among the individual qubits. They generalize the famous spin-squeezing condition
\cite{SorensenNat01}:
\begin{equation}
N{\rm var}(J_y) < \langle \hat J_x \rangle^2 ~,
\label{eq_wineland_spin_squeezing}
\end{equation}
that detect entanglement useful for metrology beyond the standard quantum limit \cite{wineland1992}. On the other hand, the condition:
\begin{equation}
(N-1){\rm var}(J_z) < \langle\hat J_x^2 \rangle + \langle \hat J_y^2 \rangle - N/2 
\label{eq_Dicke_squeezing}
\end{equation}
detects entanglement close to Dicke states \cite{LueckePRL14}, which are common eigenstates of ${\bf J}^2$ and $\hat J_z$. In particular, Dicke states with $J_z=0$ (also called twin Fock state) are useful for metrology, and can achieve the so-called Heisenberg limit of sensitivity (see next section, and ref.~\cite{pezzeRMP} for a review). The seminal work of ref.~\cite{tothetal2009} was then extended to collective spins of generic spin-$j$ ensembles \cite{vitaglianoetal2011,vitaglianoetal2014}, and to arbitrary local observables for generic qudit subsystems \cite{vitaglianoetal2011,muller-rigat_2022}, which is especially relevant in the context of experiments with atoms carrying a high magnetic moment, such as Chromium, Ytterbium or Erbium, where the populations in many Zeeman sublevels can be controlled and measured.

As done for entanglement detection, one could ask whether nonlocality can also be concluded from collective measurements only. In the device-independent setting of a Bell test, the local observables are not specified. Instead, the only information available are probabilities or correlators, reflecting the outcome statistics for given measurements performed. For instance, expectation values of Pauli operators on spin-1/2 particles are replaced by abstract probabilities $p(a\vert x)$ of obtaining result $a=\pm 1$ given some measurement setting $x$, which here would label a direction in space. The question of existence of a classical hidden-variable model compatible with the corresponding one- and two-body correlations averaged over all permutations was first successfully addressed in ref.~\cite{turaetal2014}, where a novel many-body Bell's inequality valid for an arbitrary number of particles was reported. This turned out to be especially suited to spin-squeezed states and, as further discussed in Section~\ref{sec4}, it lead to the first experimental detection of Bell correlations in atomic ensembles composed by up to half million particles \cite{schmiedetal2016,engelsen_bell_2017}. These seminal results have since been extended in several directions, including the construction of witnesses for Bell correlation depth \cite{BaccariDepth}, the investigation of statistical loopholes and many-setting witnesses \cite{Wagner17}, the formulation of device-independent witnesses of entanglement depth \cite{AloyPRL19}, and algorithms to infer permutationally-invariant Bell's inequalities directly from experimental data \cite{fadelT2017,mullerrigat2021} with a computational complexity independent of system size. This lead to the discovery of new families of Bell's inequalities, tailored to classes of states beyond spin-squeezed states of spin-1/2 ensembles, such as many-body singlets for arbitrary spin-$j$ ensembles \cite{mullerrigat2021}.

Both in the context of multipartite entanglement witnesses and Bell inequalities, it remains an important open issue to generalize known approaches to the case where higher-order moments of collective observables are available.

\subsection{Detecting multipartite correlations useful for quantum metrology}

A very important connection exists between multipartite entanglement and quantum metrology \cite{pezzeRMP}. Consider a parameter $\theta$ which has to be estimated from measurements on a quantum system, whose quantum state is the density matrix $\rho(\theta)$, encoding the unknown parameter $\theta$. The ultimate precision which can be achieved on the estimation of $\theta$ is given by the so-called quantum Fisher information (QFI), which, roughly speaking, estimates the rate at which $\rho(\theta)$ changes with $\theta$, using a suitable metric for quantifying the distance between $\rho(\theta)$ and $\rho(\theta+d\theta)$ (namely: the Bure's distance). The minimal variance on the estimated value of $\theta$ is indeed bounded by ${\rm var}(\theta) \ge 1/(\nu F_Q(\rho))$, with $F_Q$ the QFI and $\nu$ the number of independent measurements. Finding and preparing quantum states with maximal QFI in a given scenario is hence the key to improve the sensitivity of a quantum measurement device. 

Let us consider a situation akin to interferometry, where the parameter $\theta$ is encoded in the initial state $\rho$ as $\rho(\theta)=e^{-i\theta\hat O} \rho e^{i\theta\hat O}$. Here, $\hat O= \sum_{i=1}^N \hat O_i$ is some additive observable, with $\hat O_i$ acting only on subsystem $i$. In typical applications, $\theta$ could be proportional to the magnetic field intensity, and $\hat O$ to the magnetic moment of the system along the direction of the magnetic field. It can be proved that in this case, if the state is fully separable, then the QFI is bounded by $F_Q(\hat O) \le \sum_{i=1}^N (o_i^{\rm max} - o_i^{\rm min})^2$, with $o_i^{\rm max/min}$ the max/min eigenvalue of $\hat O_i$ \cite{pezzeRMP}. This result implies that, for fully-separable states, the QFI scales at most linearly with the system size $N$; hence the ultimate variance on the estimation of $\theta$ scales no faster than $1/N$, corresponding to the so-called standard quantum limit. However, since for pure states it holds $F_Q(\hat O) = 4{\rm var}(\hat O)$, it is possible to find entangled states for which $\left[\sum_{i=1}^N (o_i^{\rm max} - o_i^{\rm min})\right]^2 \sim N^2$; hence the ultimate variance for parameter estimation can scale as fast as $1/N^2$, corresponding to the so-called Heisenberg limit. Reaching this ultimate bound as allowed by the laws of quantum mechanics requires however to prepare genuine multipartite entangled states (the prominent example being GHZ states), which is extremely challenging for more than a few tens of particles, because those states are extremely prone to decoherence mechanisms. However, in-between the extremes of fully-separable states, and genuine multipartite entangled states, it can be proved that there are $k$-partite entangled states allowing one to reach a QFI scaling as $F_Q(\hat O) \sim kN$, which already overcomes the standard quantum limit \cite{hyllusetal2012,toth2012,pezzeRMP,renetal2021}. In the following, for the sake of concreteness, we focus on the relevant case where $\hat O_i$ are spin-1/2 operator, so that $(o_i^{\rm max} - o_i^{\rm min})^2=1$, and for all fully-separable states the QFI obeys $F_Q(\sum_i \hat O_i) \le N$. It should be emphasized here that there are $k$-partite entangled states having $F_Q \le N$, since not all entangled states are useful for metrology. In this sense, observing $F_Q > N$ is only a sufficient condition for entanglement. Extensions to the QFI in the form of a QFI matrix have also been proposed to the simultaneous estimation of multiple parameters and characterisation of entanglement in multi-mode systems \cite{gessneretal2019,fadelYadinetal2022}.

A large effort has then been devoted to identify families of many-body entangled states useful for metrology, and concrete scenarios to prepare and measure them. The most prominent example are spin-squeezed states, twin Fock states, and GHZ states, as prepared and manipulated in many experiments involving atomic ensembles or trapped ions. For a review of these works we refer the reader to ref.~\cite{pezzeRMP}, while our goal here will rather be to illustrate how such metrological concepts gave rise to new methods to probe quantum correlations in many-body systems. The basic idea is simple: given the fact that $F_Q>N$ implies entanglement, any measurement which allows one to at least lower-bound the QFI to a value larger than $N$ would detect multipartite entanglement (and, in particular, entanglement useful for quantum metrology) via the results of refs.~\cite{hyllusetal2012,toth2012,renetal2021}. It is worth emphasizing that this approach is especially suited to non-Gaussian states, where simple measurements of low-order moments are insufficient to observe entanglement. We discuss in the following two approaches in this spirit.

First, a rather straightforward possibility is to use the definition of the QFI, namely its connection to the ultimate precision achievable in a metrology scenario. If one measures some observable $\hat A$, its mean value $\langle \hat A \rangle(\theta) = \text{Tr}[\rho(\theta) \hat A]$ depends in general on the parameter $\theta$ to be estimated. We have $i\partial_\theta \langle \hat A \rangle = \langle[\hat A, \hat O]\rangle$, which should be compared to the variance ${\rm var}(\hat A)$. Hence, the signal-to-noise ratio is given by $[\partial_\theta \langle \hat A \rangle]^2 / {\rm var}(\hat A) \le F_Q(\hat O)$ by definition of the QFI. Therefore, from the measurement of $\langle[\hat A, \hat O]\rangle$ (either directly, or by implementing the unitary transformation) and ${\rm var}(\hat A)$ for some observable $\hat A$, one obtains a valid lower-bound to the QFI associated to unitary transformations generated by $\hat O$. The problem is then to find the most suitable observable $\hat A$ given the quantum state at hand $\hat \rho$, the generator of the transformation $\hat O$, and also given the experimental possibilities in implementing the relevant unitary transformation and in measuring the required observable $\hat A$. There is no unique solution to this problem, yet some general approaches have been proposed in this direction \cite{gessneretal2019}. As particularly relevant examples, we mention explicitly spin squeezed states, twin Fock states and GHZ states for $N$ spin-1/2 particles (see \cite{pezzeRMP} for a review). For spin-squeezed states having mean-spin along $x$ and minimal variance along $y$, one takes $\hat A=\hat J_x$ and $\hat O=\hat J_z$. For twin Fock states $\vert TF\rangle$, which are common eigenstates of ${\bf J}^2 \vert TF\rangle = (N/2)(N/2+1) \vert TF\rangle$ and $\hat J_z \vert TF\rangle=0\vert TF\rangle$, one takes $\hat A = \hat J_z^2$ and $\hat O=\hat J_x$ or $\hat J_y$. For GHZ states, defined as $|GHZ\rangle=(|\uparrow\rangle^{\otimes N} + |\downarrow\rangle^{\otimes N})/\sqrt2$, one takes $\hat A = \prod_{i=1}^N \hat \sigma_i^x$ and $\hat O = \hat J_z$.

A second more indirect approach has been developed in the context of thermal-equilibrium states, identifying signatures of many-body quantum coherence and entanglement in linear-response functions \cite{haukeetal2016,frerotR2016}. Such methods are intrinsically scalable to many-body systems for which such linear-response properties can be evaluated, both in theory and in experiments. In ref.~\cite{haukeetal2016}, the QFI has been expressed as a function of dynamical susceptibilities, which characterize the linear response of the system to time-dependent perturbations. On the other hand, in ref.~\cite{frerotR2016} the so-called quantum variance was defined as the difference between the variance and the static susceptibility (multiplied by the temperature), which are equal in classical systems. The quantum variance was shown to be a tight lower-bound to the QFI, while requiring only to probe the linear response to static perturbation, which is much less demanding to evaluate than dynamical response functions. These works establish a deep link between quantum statistical physics and quantum entanglement useful for metrology, and offer scalable tools to probe quantum entanglement in many-body systems at thermal equilibrium.

\subsection{Insight into quantum phase transitions}

The traditional Landau theory of phase transitions relies on the notion of local order parameters which define phases of matter. In an ordered phase, the order parameter displays long-range correlations, which start to build up already in a disordered phase upon approaching the phase transition. This picture is valid both for classical and quantum standard phase transition. However, fundamental discoveries over the last thirty years have brought a new picture as well as a novel paradigm to classify phases of matter. This has concerned quantum phase transition in the first place, and then topological phase transitions. In these instances, describing correlations at a classical level only is not sufficient. Indeed, the contemporary understanding of quantum, and especially topological quantum phase transitions, is tightly related to the understanding of quantum entanglement properties at the many-body level. On the one hand, standard quantum phases are well described by short-range entangled states, while the range of entanglement grows only at criticality. The situation is much more complex with topological quantum phases, where characterization of the phase is related to topological entanglement entropy or entanglement spectrum  \cite{levinW2006,wenRMP}, and the first remarkable example concerns symmetry protected topological phases in one dimension. For a recent overview of the role of quantum correlations in  topological phases, the reader should  consult the book by X.-G.~Wen and collaborators \cite{Wen-new}.

Quantum phase transitions (QPT), as opposed to thermal phase transition, are driven by quantum fluctuations in the ground state of many-body systems when varying an external parameter \cite{sachdev2011quantum}. At the critical point of a second-order QPT, the coherence length for quantum correlations diverges, potentially leading to rich entanglement patterns within the many-body system. A first example of this phenomenon was predicted in an Ising model with all-to-all interactions, described by the Hamiltonian $\hat H = -(J/N)\hat J_z^2 - B\hat J_x$ with $N$ the number of qubits. It was predicted that in the ground state, at the critical point $J=B$, scalable spin-squeezing occurs (namely, $\langle \hat J^2_y \rangle \sim N^{2/3}$ and $\langle \hat J_x \rangle \sim N$, so that the spin-squeezing parameter is $\xi_R^2 =N\langle \hat J_y^2 \rangle / \langle \hat J_x\rangle^2 \sim 1/N^{1/3}$, scaling to zero in the thermodynamic limit) \cite{dusuelV2004} (see ref.~\cite{Makhalovetal2019} for a recent experimental manifestation of the phenomenon of quantum-critical spin squeezing inside individual Dysprosium atoms, whose $J=8$ electronic spin is formally equivalent to $N=16$ spin-1/2). The usefulness of QPT for quantum metrology was later shown to be a general result by analyzing the scaling properties of the quantum Fisher information \cite{haukeetal2016}, which at the QPT displays a divergence that is robust to thermal noise \cite{FrerotR_2019,Gabbrielli_2018}. This remarkable result, implying that the kind of multipartite entanglement produced around quantum critical points is a resource for quantum metrology, gave birth to the notion of quantum critical metrology \cite{frerotR2018}, and was exemplified by the theoretical prediction of robust spin-squeezed states at the QPT of quantum Ising models, beyond the all-to-all interacting model initially considered in ref.~\cite{dusuelV2004}.

Given that quantum entanglement at all scales is stabilized in the vicinity of QPT, it is natural to investigate also there the violation of multipartite Bell's inequalities, and their robustness at finite temperatures. First studies in that direction were proposed in refs.~\cite{piga_bell_2019,frerotR2021_bell}, and the robustness of Bell correlations at finite temperatures in a spin model with all-to-all interactions was studied in \cite{Fadel2018bellcorrelations}. We notice that the connection between quantum phase transitions and quantum correlations in developed in several monograph handbooks, such as ref.~\cite{LSA2017} by one of us as well as the more focused book of ref.~\cite{Wen-new} by X.-G.~Wen and collaborators.

\subsection{Detecting quantum correlations between two many-body systems}
So far we have reviewed methods to probe multipartite correlations between the individual constituents (typically qubits) of a many-body system. Here, instead, we would like to mention approaches to probe quantum correlations between two many-body systems, $A$ and $B$. These may be two disjoint partitions of a single many-body system; or they can be two spatially-separated systems. 

An important concept in bipartite scenarios is that of entanglement entropy, which is defined as the (von Neumann) entropy of a subsystem $S(\hat \rho_A)=-{\rm Tr}(\hat \rho_A \log \hat \rho_A)$ with $\hat \rho_A={\rm Tr}_B(\hat \rho_{AB})$ the reduced state of $A$ subsystem. When the global $AB$ system is in a pure state, then any non-zero entropy for $A$ subsystem is a proof of entanglement. More generally, the condition $S(\hat \rho_A) > S(\hat \rho_{AB})$ implies the $A$ and $B$ are entangled \cite{horodecki_RMP_2009}. In experiments, one rather reconstructs the so-called Rényi entropy, defined as $-\log {\rm Tr}(\hat \rho_A^2)$, and which has similar properties to the von Neumann entropy regarding its scaling with system size, or its ability to detect, \eg  topological order and quantum phase transitions. Rényi entropy may be reconstructed by probing quantum interferences between two copies of the many-body system under study, as implemented in several experiments in the recent years \cite{Islam2015,KaufmanSci16,Lukin_2019,Bluvstein_2022}. The basic idea is that ${\rm Tr}(\hat \rho^2)$ can be written as the mean value of the SWAP operator on $\hat \rho \otimes \hat \rho$, which in turn can be implemented as a beam-splitter transformation between two copies of the many-body state $\hat\rho$. The quantity ${\rm Tr}(\hat \rho_A^2)$ is then accessed by data post-processing, where subsystems $A$ and $B$ can be arbitrarily defined. More recently, an alternative approach requiring a single copy and randomized measurement schemes has been developed \cite{Brydges_2019}. This seminal work has since been extended to a rich toolbox for characterizing quantum many-body systems, for which a detailed review has been recently published \cite{randomized_review}. Notice that approaches based on reconstructing subsystem entropies to probe bipartite entanglement, either using two copies or using randomized measurements, are intrinsically limited to modest size systems (at most a few tens of qubits for the most advanced randomized measurements approaches), as the required number of measurements scales exponentially with the number of qubits. Moreover, one has typically to assume, or experimentally verify, that the system is (at least close to being) in a pure state with $\text{Tr}[\hat \rho^2]\approx 1$. On the other hand, since the choice of the $AB$ bipartition where to probe entanglement can be defined a posteriori during data analysis, this approach has the advantage of allowing one to explore entanglement for any possible bipartition of the system with the same experimental setup and collection of measurements.

Complementary approaches rely on low-order moments of collective observables measured locally on the subsystems. In this context, also motivated by developments in cold atoms experiments, bipartite entanglement witnesses from measurement of local collective spins have been proposed, tailored to split squeezed Bose-Einstein condensates \cite{Jing2019}, and split Dicke states \cite{vitaglianoFadeletal2021}. The same witnesses can be used to bound entanglement measures, thus giving a proper quantification of the ``strength'' of correlations present in the system \cite{MorrisPRX20,FadelPRL21}. Moreover, a slight modification of these bipartite witnesses allows one to detect also correlations stronger than entanglement, namely EPR-steering \cite{Jing2019,vitaglianoFadeletal2021}. Further studies resulted in even better criteria for the detection of this stronger type of correlations, which are based on the measurement of generalised spin-squeezing parameters \cite{Jiajie21} or of the Fisher information \cite{YadinNatCom21}, and for this reason also unveiling that EPR-steering can be a resource for quantum metrology tasks.
On the other hand, the search for Bell inequalities suited to these scenarios turned out extremely challenging. Taken into account known no-go theorems \cite{KaszliPRL2011}, the detection of nonlocality between a pair of many-body systems seems to require nonlinear measurement schemes \cite{KitzingerPRA21}.

To conclude, let us mention that data-driven methods initially devised for probing multipartite entanglement can be adapted to probe also bipartite entanglement, and to find new witnesses \cite{frerotetal2022}. Moreover, first steps towards quantifying entanglement dimensionality in such bipartite many-body systems from local collective measurements have been reported in \cite{Liu21}.

\section{Experimental investigations}\label{sec4}
In this Section, we will present results on the experimental detection of quantum correlations in multipartite systems. These have been achieved in a variety of physical platforms, and considering different physical degrees of freedom. The latter can be broadly classified into discrete variables, such as atomic spins or light's orbital angular momentum, or continuous variables, such as phase-space quadrature of mechanical oscillators or of electromagnetic modes. 

We will first describe the advantages and limitations of each different experimental platform, with a focus on the available tools for quantum state preparation and measurements. Then, we will give an overview over the major progresses in multipartite entanglement detection achieved over the last decade, and discuss the new aspects which could be investigated with current theoretical and experimental capabilities. Finally, we present the main technical barriers to further progress, as well as conceptual open problems.

%\subsection{Platforms}

\subsection{Atomic ensembles}

\begin{figure*}[ht]
    \centering
  \includegraphics[width=0.8\textwidth]{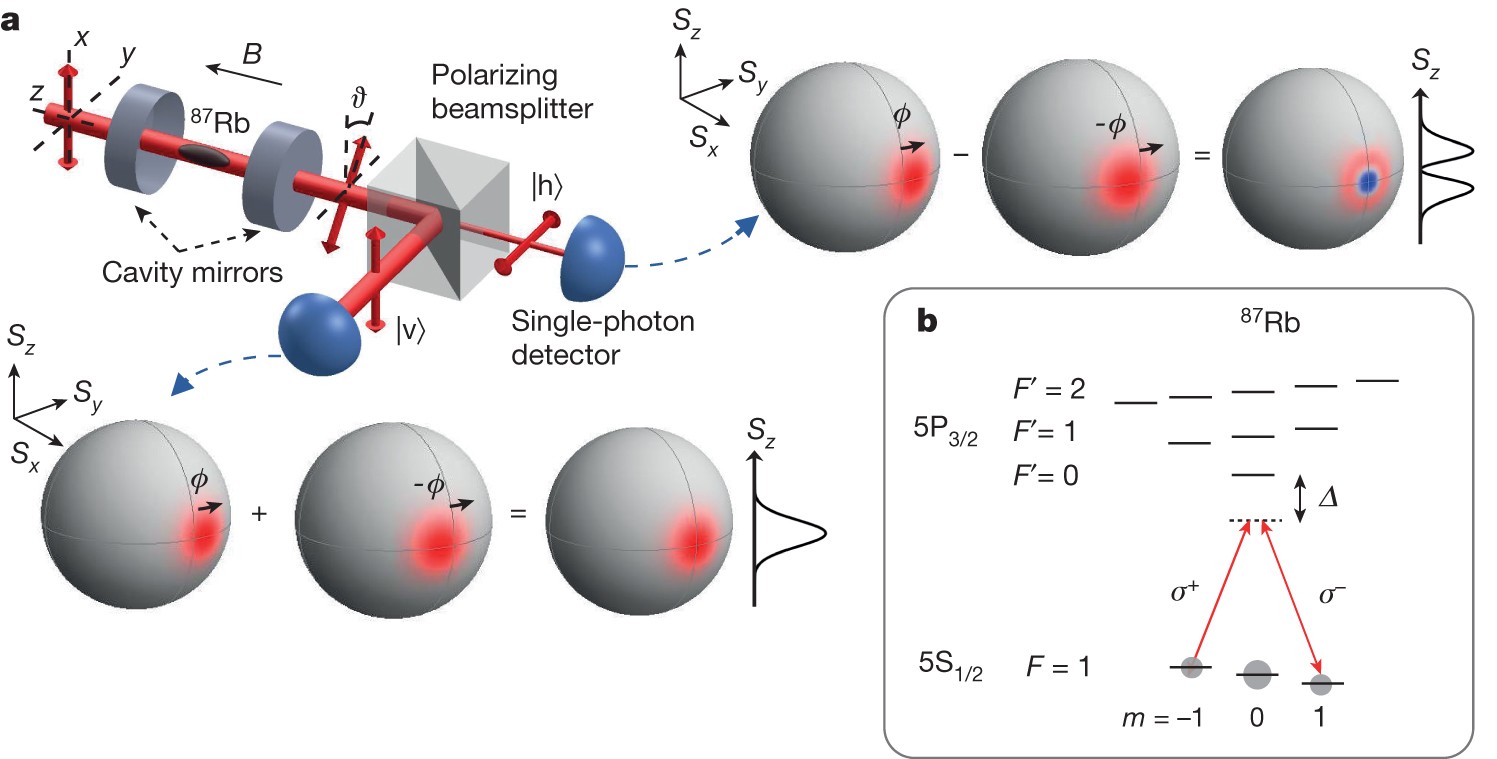}
    \caption{\textbf{Heralded entanglement generation in atomic ensembles by single-photon detection.} Panel \textbf{a}: Incident vertically polarized light (top left) experiences weak polarization rotation due to atomic quantum noise, and the detection of a horizontally polarized transmitted photon heralds an entangled state of collective atomic spin. An optical resonator formed by two mirrors enhances the polarization rotation and the heralding probability. The Bloch spheres show the ideal Wigner distribution functions for the collective spin upon registering the corresponding single-photon detection events. Panel \textbf{b}: $^{87}$Rb level structure. Atoms in the $5S_{1/2}$, $F=1$ hyperfine manifold are coupled to the excited $5P_{3/2}$ manifold via linearly polarized light, decomposed into two circular polarization components (indicated by red arrows) that interact with the atomic ground-state populations. The outgoing polarization state of the light reflects the quantum fluctuations between the $\vert 5S_{1/2} F=1, m=\pm 1 \rangle$ magnetic sublevels. Reproduced from ref.~\cite{McConnell15}.}
    \label{fig:atom1}
\end{figure*}

Cold and ultracold atomic ensembles are ideal platforms to investigate quantum correlations among a number of particles ranging from $\sim 10^2$ to $\sim 10^6$ with collective measurements. Typically, alkali atoms are laser-cooled and then confined in magnetic or optical traps. Additional cooling mechanisms allow one to decrease the atoms kinetic energy even further, which can result in Bose-Einstein condensation (BEC). Interesting nonclassical states are prepared in the internal (i.e. spin) degrees of freedom by controlling inter-particle collisions or the collective atom-light interaction with a travelling optical field or cavity mode. These interactions give rise to an effective non-linear dynamics that creates correlations among the particles, or it allows for measurement-based state preparation (see Fig.~\ref{fig:atom1}). Neglecting the external (i.e. motional) degrees of freedom, the quantum state of the ensemble can be described by a collective spin state, that can be conveniently depicted on a generalized Bloch sphere. Spin rotations are implemented via Rabi pulses, that collectively address the entire ensemble. The quantum state can then be analyzed by performing collective spin measurements along different axes, either via Stern-Gerlach-type measurements, or via Faraday interaction with a detuned optical beam.

Advantages of these experimental platforms consist in the possibility of preparing and manipulating multipartite quantum states in a collective fashion, and thus applicable to a number of particles spanning several orders of magnitude. The latter is usually limited by interaction effects, trap depth and inhomogeneity. 

On the other hand, the collective interaction results in the impossibility of addressing independently each of the constituents, which in the case of BECs is even impossible in principle, as all particles are fundamentally indistinguishable. When dealing with large number of particles it also becomes difficult to suppress particle losses, a phenomenon resulting in decoherence and phase noise, that ultimately hinders the preparation and manipulation of highly entangled states such as $N00N$ states. Moreover, it is extremely challenging to implement measurements with single particle resolution, meaning that the final state is typically reconstructed with additional noise.

Spin squeezing in atomic ensembles was originally observed in $10^{7}$ Cesium atoms resonantly pumped with squeezed light \cite{HaldPRL99}. Reduction in the spin projection noise compared to uncorrelated atoms was detected through polarization analysis of an optical probe beam, and allowed to conclude the presence of entanglement. 

Spin-squeezing in BECs was originally observed in \cite{EsteveNat08} through Eq.~\eqref{eq_wineland_spin_squeezing}, where the spin degree of freedom was defined from the occupation number of two adjacent trap sites. This was later extended to a collective spin defined from two internal (hyperfine) ${}^{87}\text{Rb}$ states in \cite{GrossNat2010,RiedelNat10}, where a quantification of the number of entangled particles was also given. In these two experiments the generation of multi-particle entanglement was based on controlling elastic collisional interactions through a Feshbach resonance \cite{GrossNat2010}, or through a state-dependent potential \cite{RiedelNat10}.

Apart from spin-squeezed BECs, a quantification of entanglement by means of the so-called entanglement depth, defined as the number of particles in the largest nonseparable subset, has been also investigated for BECs in Dicke states in \cite{LueckePRL14} through the generalized spin squeezing criterion Eq.~\eqref{eq_Dicke_squeezing}. Dicke states can be prepared through spin-exchange collisions in the $F=1$ manifold of ${}^{87}\text{Rb}$ atoms, a process resembling optical parametric down-conversion. Operating in this three-level manifold also allowed for the preparation of spin-nematic squeezed states \cite{hamleyetal2012}, which are described in terms of $\text{SU}(3)$ dipole-quadrupole operators, of $N=100$ atoms spin-singlet states \cite{evrardetal2021}, and of $N=10^4$ atoms twin-Fock states \cite{luoetal2017}.

Spin squeezed states have also been realized in atomic gases through quantum non-demolition (QND) interaction of the spin state with an off-resonant light mode, either confined in an optical cavity \cite{schleiersmithetal2010,Leroux2010,chenetal2011} or in the form of a travelling field \cite{sewelletal2012}. These measurement-based state preparation techniques can be used to prepare also states different from spin-squeezed states, and they apply not only to BECs \cite{HaasSCI14}, but also to cold ensembles \cite{McConnell15} and hot gases \cite{KongNatCom20}. In  ref.~\cite{HaasSCI14} small BECs ($N=41,23,12$) were strongly coupled to an optical cavity mode, which allowed for preparation and measurement of $W$ states. From a partial tomographic reconstruction of the atoms state, multipartite entanglement was investigated concluding the presence of at least 13 entangled atoms. Afterwards, ref.~\cite{McConnell15} demonstrated the preparation of $3000$ cold atoms in an entangled state with negative Wigner function, heralded by single-photon detection, see Fig.~\ref{fig:atom1}. More recently, ref.~\cite{KongNatCom20} showed the preparation of singlet-type entangled states of $10^{13}$ atoms at $T=463\,\text{K}$. A complementary approach to measurement-based state preparation involves exploiting quantum Zeno dynamics, where a frequent enough measurement can confine the system dynamics to a small subspace. This has allowed for the preparation of nonclassical states in ensembles of 35 atoms, for which entanglement among at least $\sim 8$ particles was concluded through full tomography in the symmetric subspace \cite{Barontini15}.

Besides multipartite entanglement detection and quantification based on spin-squeezing coefficients, BECs prepared in non-classical spin states were used also for demonstrating entanglement detection by means on the Fisher information \cite{StrobelSci14}, and entanglement quantification in terms of entanglement monotones \cite{FadelPRL21}.

The possibility of detecting multipartite quantum correlation stronger than entanglement was suggested by the formulation of Bell inequalities involving only one- and two-body correlators, and thus of practical experimental implementation \cite{turaetal2014}. This allowed for the demonstration of multipartite Bell correlations in spin-squeezed BECs \cite{schmiedetal2016}, and later in cold atomic ensembles \cite{engelsen_bell_2017}. A further development of these theoretical tools made possible to use these experimental data also for the quantification of Bell correlation depth \cite{BaccariDepth}, and the device-independent detection of entanglement depth \cite{AloyPRL19}.

The experimental results mentioned so far in this section rely on performing collective measurements on atomic ensembles, where the constituent particles are not resolved. For this reason, interest was aroused towards the possibility of ``extracting'' such correlations in individually addressable subsystems. Experiments implementing this idea allowed for the demonstration of entanglement between two spatial regions in BECs \cite{Fadeletal2018,Langeetal2018,kunkeletal2018}, and even of EPR-steering \cite{Fadeletal2018,kunkeletal2018}. This method was also used to investigate experimentally the concept of indistinguishable-particle entanglement and its quantification in refs.~\cite{MorrisPRX20,FadelPRL21}, and performing local measurements on more than two partitions was used to demonstrate multipartite entanglement structure in refs.~\cite{kunkeletal2018,KunkelPRL22}.

\subsection{Atoms in lattices}

\begin{figure*}
    \centering  \includegraphics[width=0.9\textwidth]{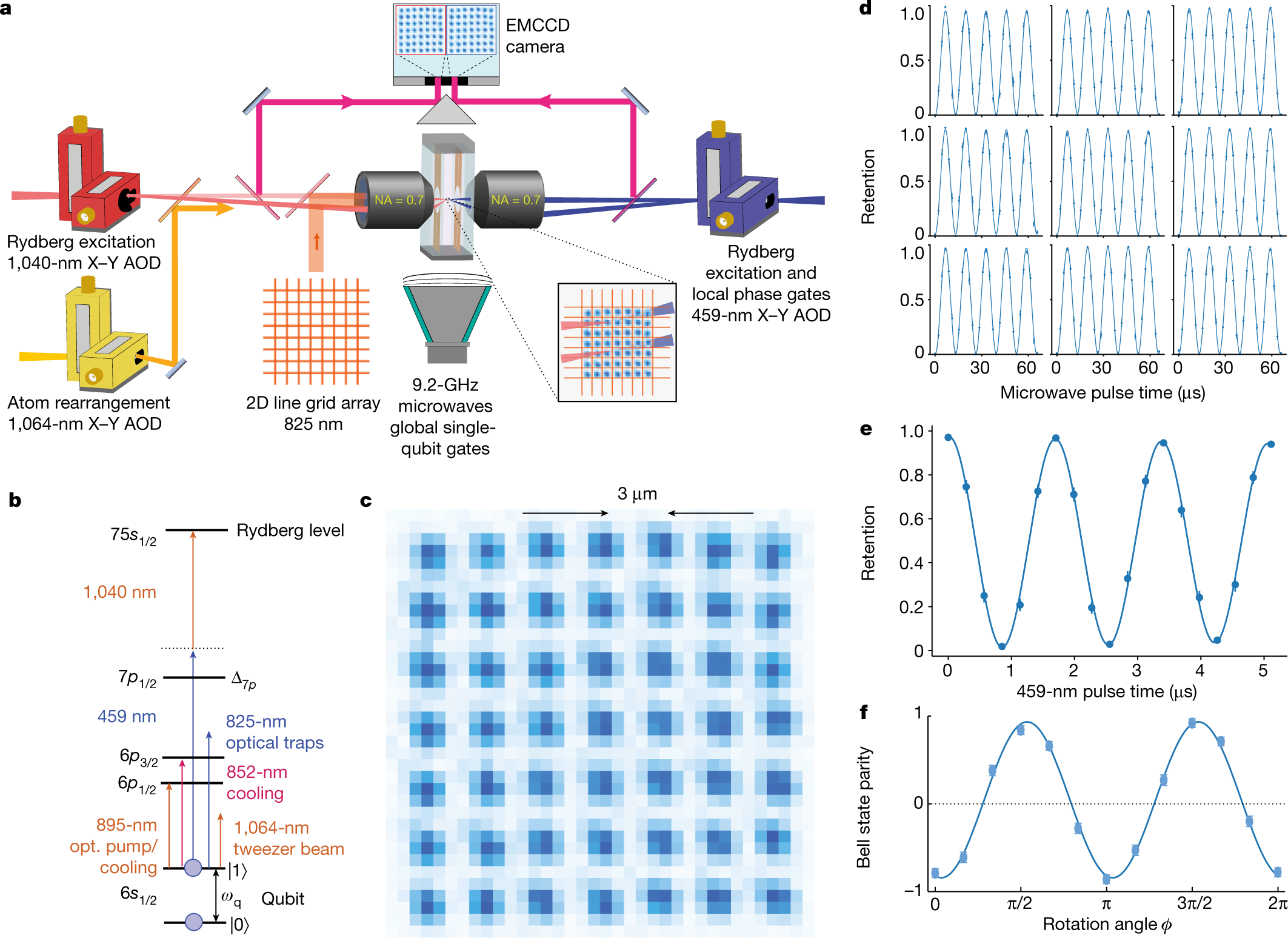}
    \caption{\textbf{Lattice of Rydberg atoms for quantum computing.} Panel \textbf{a}: Experimental layout for trapping and addressing atomic qubits. Atoms are trapped in a blue-detuned line grid array, which is imaged onto the atom-trapping region with a lens. Atom occupation is determined by collecting atomic fluorescence using lenses at opposite faces of the cell and imaging the light onto two separate regions of an EMCCD camera. A \unit{1064}{nm} tweezer beam is used to rearrange atoms into desired sites for circuit operation. Circuits are decomposed into a universal gate set consisting of global rotations about an axis in the $xy$ plane driven by microwaves, local rotations driven by the \unit{459}{nm} beam and CZ entangling gates using simultaneous Rydberg excitation of atom pairs by the \unit{459}{nm} and \unit{1040}{nm} beams. Panel \textbf{b}: Cesium atom level diagram, and wavelengths used for cooling, trapping and qubit control. Panel \textbf{c}: Averaged atomic fluorescence image of the 49-site array with spacing \unit{3}{\mu m}. Panel \textbf{d}: Global microwave Rabi rotations on a block of nine qubits at \unit{76.5}{kHz}. Panel \textbf{e}: A Ramsey experiment with microwave $\pi/2$ pulses and the focused \unit{459}{nm} beam providing a rotation on a single site. Panel \textbf{f}: Parity oscillation of a two-qubit Bell state created using a CZ gate. The measured and uncorrected Bell state fidelity was $92.7(1.3)\%$ for an optimized qubit pair (about $95.5\%$ corrected for state preparation and measurement (SPAM) errors), with the average for all connected qubit pairs used in circuits being $90\%$ (about $92.5\%$ SPAM corrected). Reproduced from ref.~\cite{Graham_2022}. }
    \label{fig:atom2}
\end{figure*}

In the previous section we have discussed ensembles of neutral atoms confined in a single trapping potential. This has the advantage of allowing for the collective manipulation of a large number of particles, but with the drawback of loosing independent addressing of the individual atoms. Experimental platforms where the latter feature is available are artificial lattices equipped with quantum gas microscopes. In this case, neutral atoms are confined in an array of potential minima, generated by interfering optical beams, optical tweezers, or by static magnetic fields. A careful tuning of parameters such as the depth of the traps, the tunneling rate between adjacent minima, and the atoms collisions allow for the preparation of interesting quantum many-body states. For example, these tools allow for the preparation of lattices with single occupation per site (Mott insulator regime), which is often the starting point for a number of experiments. Then, by further tuning the system's parameter, it is possible to implement paradigmatic Hamiltonians (such as Bose-Hubbard, Fermi-Hubbard, ...). Some remarkable experiments involve lattices inside optical cavities, that are used to mediate long-range interactions between sites that are not adjacent, or arrays of Rydberg atoms (see Fig.~\ref{fig:atom2}), that can be controlled with exquisite precision.
Multipartite quantum states of interest can be encoded in the lattice occupation basis, or in the atoms internal (spin) degree of freedom. These states are reconstructed by performing measurements in the appropriate basis, with the help of quantum gas microscopes to resolve individual sites. 

Two-components BECs in spin-dependent superlattices have been used to prepare ensembles of Bell states with fidelity $\gtrsim 80\%$ \cite{Dai16,Yang_2020}, showing a violation of the CHSH inequality $S=2.21\pm 0.08$ \cite{Dai16}. Recently, this platform was used for the preparation of 1D arrays of 10 entangled atoms, and 2D plaquettes of $2\times 2$ and $2 \times 4$ entangled atoms \cite{zhangetal2022}.

Simulation of Bose-Hubbard hamiltonians resulted in the preparation of multipartite states whose entanglement in the occupation basis was confirmed through many-body interference, and quantified through the Rényi entropy \cite{Islam2015,KaufmanSci16}. Entropy and density-density correlations have also been adopted to investigate the evolution and scaling of entanglement during many-body out-of-equilibrium dynamics \cite{Lukin_2019,Rispoli_2019}. Alternatively, entanglement between lattice sites has been quantified by bounding LOCC monotones from measurement of the atomic density distribution after time-of-flight \cite{CramerEtAl2013}. In one-dimensional Bose-Hubbard chains, entanglement was observed in the internal (spin) degree of freedom in \cite{fukuharaetal2015}.

Rydberg-atom arrays have been used to prepare GHZ states of up to six (\cite{Graham_2022}, see Fig.~\ref{fig:atom2}), and 20 qubits \cite{omran_generation_2019}. More recently, the same platform allowed for the preparation of entangled graph states, such as cluster states and a seven-qubit Steane code state, and for measuring the time evolution of entanglement entropy and Rényi entanglement entropy after a quench \cite{Bluvstein_2022}.

\subsection{Trapped ions}

\begin{figure}
    \centering
\includegraphics[width=\columnwidth]{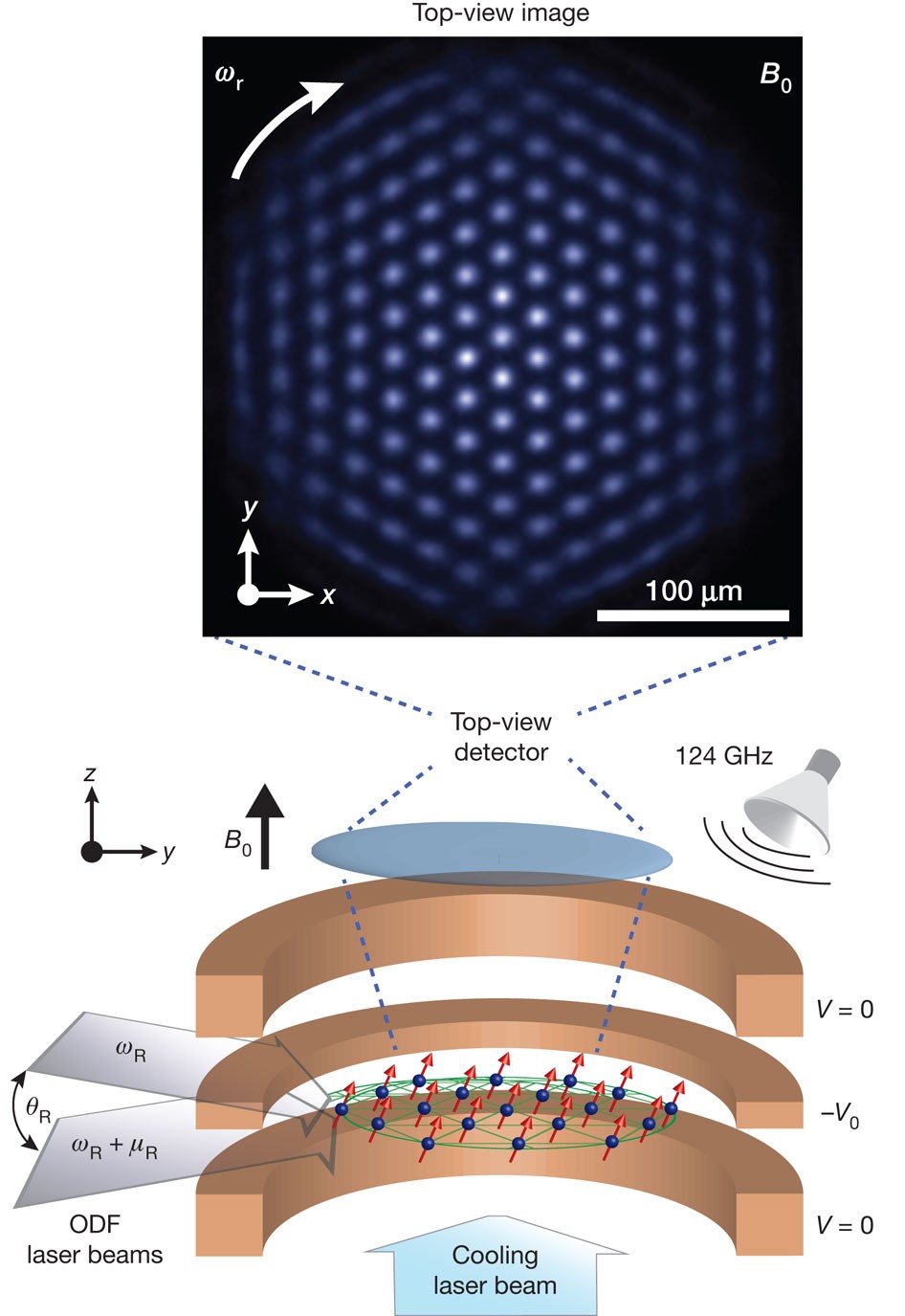}
    \caption{\textbf{Penning trap for two-dimensional ion crystal.} Each qubit is the valence-electron spin of a ${}^9 \text{Be}^+$ ion. Bottom: a Penning trap confines ions using a combination of static electric and magnetic fields. The trap parameters are configured such that laser-cooled ions form a triangular 2D crystal. A general spin–spin interaction is generated by a spin-dependent excitation of the transverse (along $z$) motional modes of the ion crystal. This coupling is implemented using an optical dipole force produced by a pair of angled off-resonance laser beams (left side). Microwaves at \unit{124}{GHz} permit global spin rotations. Top: a representative top-view resonance fluorescence image showing the centre region of an ion crystal captured in the ions rest frame; in the laboratory frame, the ions rotate at $\omega_r=2\pi\,43.8\,$kHz. Fluorescence is an indication of the qubit spin state (up=bright; down=dark); here, the ions are in the up state. The lattice constant is $\approx 20\,\mu$m. Reproduced from ref.~\cite{Britton12}.}
    \label{fig:ion2}
\end{figure}

Originally conceived for precision spectroscopy and atomic clocks, ion trapping became of crucial importance for quantum computations and simulations. Charged particles can be conveniently confined using modulated electric fields (Paul traps) or with a combination of static electric and magnetic fields (Penning traps). In the first case, around $10-100$ ions are typically confined to form a string-like configuration, which results from the combination of trapping potential and Coulomb repulsion. In the second case, hundreds of atoms are typically confined to form a membrane-like configuration, that rotates around the static trapping magnetic field at the Cyclotron frequency, see Fig.~\ref{fig:ion2}. In these platforms, the deep trapping potentials, in combination with laser-cooling techniques, allow for a long storage of the ions (up to months). Moreover, laser or electromagnetic pulses enable the manipulation of both external and internal degrees of freedom with extraordinary high fidelity. In the case of Paul traps this manipulation can address each single ion, while the rotation of the ion crystal in a Penning trap makes single-particle addressability significantly more complicated. For this reason, in the latter platform it is common to implement collective manipulations, similarly as for the case of atomic ensembles. By performing fluorescence measurements, it is possible to take screenshots of the ions internal degree of freedom, and thus perform state tomography. 

Compared to atomic ensembles, while ion arrays in Paul traps do not suffer from particle losses and allow one for single-particle addressability and detection, they are currently limited to operate with at most a few tens of ions. One of the difficulties comes from the fact that the implementation of mutliqubit gates, and thus of engineered interactions between particles, makes use of the motional degree of freedom. If the ion crystal is composed of a large number of particles, the motional eigenmodes are close in frequency, and thus difficult to address individually, which decreases the gate fidelities.

In linear Paul traps, GHZ states of 6 and 14 ions have been prepared in refs.~\cite{Leibfried05} and \cite{MonzPRL11} respectively, and detected through entanglement witnesses. W states of up to 8 ions have been prepared in ref.~\cite{Haeffner05}, and detected by both full state tomography and generalized spin-squeezing inequalities \cite{KorbiczPRA06}. A fully-controllable 20 qubit register was used to study the out-of-equilibrium dynamics of an Ising-type Hamiltonian in ref.~\cite{Friis_2018}, where genuine multipartite entanglement was shown to appear in groups of up to 5 particles. In the same experimental setup, the use of randomized measurements enabled to estimate the second-order Rényi entropy for different partitions containing up of 10 particles, thus providing a quantification of entanglement and a characterization of its structure \cite{Brydges_2019}.
Device-independent genuine multipartite entanglement, as well as multipartite nonlocality, were reported in ref.~\cite{Barreiro13} for a GHZ state of 6 entangled ions. Graph states essential for measurement-based quantum computation have been prepared between 3 and 7 ions \cite{Lanyon14}, and used to violated suitable multipartite Bell inequalities. The same work also showed that the data from ref.~\cite{MonzPRL11} yield a violation of the Mermin-Klyshko Bell inequalities that increases exponentially with system size.

In Penning traps, see Fig.~\ref{fig:ion2}, the collective manipulation of two-dimensional arrays of $\sim 200$ ions enabled the preparation of spin-squeezed and over-squeezed states through the engineering of long-range homogeneous Ising interaction \cite{Britton12,bohnetetal2016}, and the preparation of spin-motion entangled states using spin-dependent optical dipole forces \cite{Gilmore21}. In these many-body systems that lack individual addressing of the particles, entanglement was deduced using the same tools adopted from atomic ensembles, namely the Wineland spin-squeezing coefficient and the Fisher information. In the future, technical developments could enable individual addressing of the ions in the array, thus opening up new possibilities to study quantum correlations in such setups.

\begin{figure*}[ht]
    \centering   
    \includegraphics[width=0.9\textwidth]{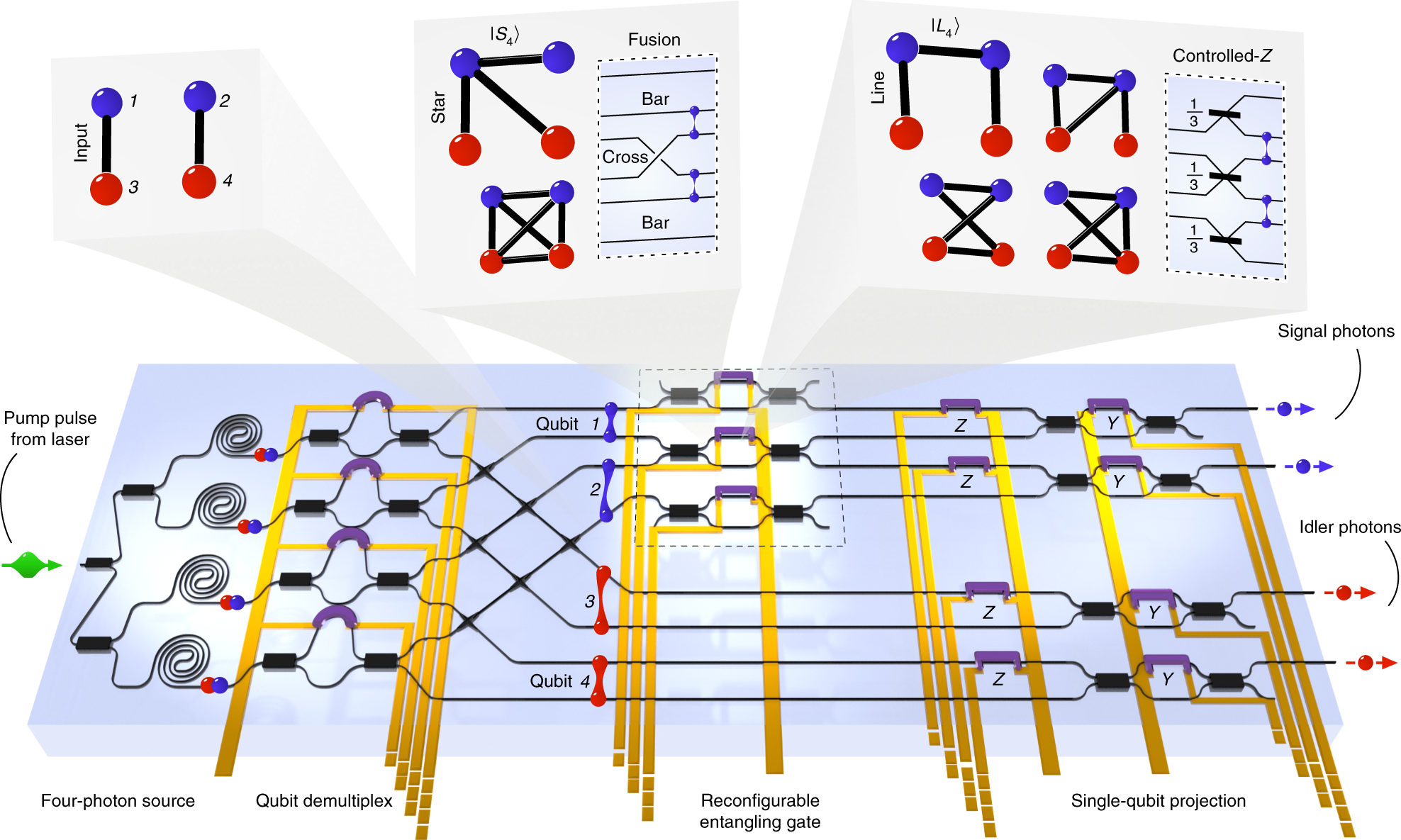}
    \caption{\textbf{Integrated photonic circuits.} Schematic of the silicon-on-insulator chip device, comprising: four telecommunications-band photon-pair sources, producing four photons in superposition; a qubit demultiplexer, which configures that superposition into a product of two Bell-pairs; a reconfigurable postselected entangling gate (R-PEG); and four single-qubit projection and analysis stages, formed of four Mach–Zehnder interferometers implementing qubit $Y$ rotations, preceded by four $Z$ rotations. Corresponding graph states are indicated above, starting with the two input Bell-pairs, and ending with either ‘star’ or ‘line’ graph states, for fusion or controlled-$Z$ R-PEG configurations, respectively. Reproduced from ref.~\cite{Adcock_2019}.}
    \label{fig:opt1}
\end{figure*}

\subsection{Optical photons}
As they interact weakly with their environment, photons are ideal systems for transmitting information, which can be encoded into several different degrees of freedom, such as frequency, polarization, path, and orbital angular momentum (OAM). This opens up also the possibility of preparing so-called hyperentangled states, where quantum correlations appear among different degrees of freedom simultaneously.

Experimental techniques to prepare entangled states of photons typically rely on non-linear crystals to realise spontaneous parametric down-conversion (SPDC) and optical parametric oscillators (OPOs).
The polarization-entangled photon pairs resulting from SPDC represent the most popular building block for the generation of multipartite entangled states. For example, one of the photons of the pair is sent to a cascade of beam-splitters that distributes it among $N$ modes, i.e. a $W$ state, while the second photon heralds the successful state preparation. On the other hand, high-oder SPDC processes can be used to generate directly 3-photon entangled states \cite{PhysRevX.10.011011}.
SPDC sources in combination with holographic spatial light modulators, namely liquid-crystal devices that can impart an arbitrary 2D phase pattern onto a photon, can be used to prepare entangled states in the OAM degree of freedom.
OPOs can be used to prepare continuous-variable single-mode squeezed states that, if made interfere in a network of beam splitters, can give rise to multimode entangled states in $xp$-quadratures.
Alternatively, pumping an OPO with a frequency comb results in continuous-variable correlations between different frequency components.

For state characterization, possible measurements that can be performed include homodyne detection of phase-space quadratures, photon-number detection, and projection into polarization/OAM basis states. An extensive review on multiphoton entanglement experiments performed until 2011 is ref.~\cite{RevModPhys.84.777}, while below we will focus on seminal results of the last decade.

The polarization DoF has been used to prepare graph states of 6 photons in ref.~\cite{Lu_2007}, and GHZ states of 8, 10, and 12 photons in refs.~\cite{huang2011experimental,YaoNatPhot12,PhysRevLett.117.210502,Chen_17,PhysRevLett.121.250505}. The preparation of multipartite entangled states using SPDC is however difficult to scale up, as it is hindered by its probabilistic nature. 

Entanglement in the time-bin degrees of freedom was demonstrated among six photons, using a mesoscopic atomic ensemble under Rydberg blockade to generate a train of correlated single-photons \cite{Yang2022}. A cavity-electrodynamics setup implemented by a single Rubidium-87 atom in a high-finesse optical cavity was used to prepare GHZ states of up to 14 photons, and cluster states of up to 12 photons \cite{Thomas_2022}.

Integrated photonic circuits, see Fig.~\ref{fig:opt1}, constitute a promising platform for optical quantum technologies. With these circuits, 4-photon GHZ states were prepared in ref.~\cite{Matthews_2009,Llewellyn2020}, 4-photon graph states in ref.~\cite{Adcock_2019}, 4-photon time-bin entangled states in ref.~\cite{doi:10.1126/science.aad8532}, and single-photon $W$ states of up to 16-modes in ref.~\cite{grafe2014chip}.

Hyperentangled 6-, 8- and 10-qubit $N00N$ states exploiting both the photons polarization and momentum degrees of freedom have been reported in ref.~\cite{Gao10}.
A 18-qubit hyperentangled GHZ state has been prepared by exploiting paths, polarization, and orbital angular momentum degrees of freedom of six photons in ref.~\cite{PhysRevLett.120.260502}.  

High-dimensional multipartite entanglement was originally demonstrated in ref.~\cite{Malik2016} for three photons, with two of the photons being in a three-dimensional space whereas the third lives in two dimensions.

Continuous-variable (i.e. $xp$-quadrature) multipartite entangled states have been investigated for cluster states involving more than ten thousand time-domain modes \cite{Yokoyama_2013}, up to 60 frequency-domain modes \cite{gerkeetal2015,PhysRevLett.112.120505,Roslund_2013,cai2017multimode}, and for three-mode non-Gaussian states based on photon-subtraction \cite{Ra_2019}.

Linear optics networks have also been used to investigate multipartite EPR steering scenarios. In particular, continuous-variable entanglement of 8 modes made possible to demonstrate up to 7-way steering and monogamy relations in a 3-partite scenario \cite{Armstrong2015}.

\subsection{Superconducting qubits}
Qubits based on Josephson junctions constitute one of the most successful platforms for quantum computations and simulations. Connecting such devices through superconducting resonators allows one to use the paradigms of cavity quantum electrodynamics to engineer interactions, and thus logic operations, between them. Currently, several tens of superconducting qubits can be independently controlled and measured with relatively high fidelity, allowing for the implementation of quantum circuits comprising hundreds of gates.  

Genuine multipartite entanglement was observed in GHZ states composed by up to 12 qubits through a fidelity-based entanglement witness \cite{PhysRevLett.122.110501}, and 20 qubits through parity oscillations \cite{doi:10.1126/science.aay0600}.

The IBM quantum computer allowed for the detection of entanglement in graph states of 16 qubits \cite{Wang_2018}, and of 65 qubits \cite{mooneyetal2021}, using negativity measurements, and of genuine multipartite entanglement for GHZ states of up to 27 qubits through parity measurements \cite{Mooney_2021}.

On the Google quantum computer, entanglement entropies in the toric code ground state on a lattice of 31 qubits were measured \cite{satzingeretal2021}.

\subsection{Condensed-matter systems}
Multipartite entanglement is expected to appear naturally in many solid-state systems, such as between the electron spins in a magnetic material. The study of quantum correlations in such systems is of great interest, especially to understand their relationship with more macroscopic (such as traditional linear-response) properties. This task is however very challenging, due to the limited control available in manipulating bulk solid-state materials. While the underlying Hamiltonian is difficult to engineer, as it depends on the lattice structure and on the atoms involved, parameters that can be easily controlled experimentally are temperature, strain, concentration of impurities, and external electromagnetic fields. Possible measurements that can be implemented rely primarily on transport, and on scattering of light or particles, such as neutrons to probe magnetic materials. Similarly to the case of atomic ensembles, single-particle addressability is practically unfeasible, so the observation of quantum correlations in solid-state materials must rely on witnesses. 

Recent experiments reconstruct the quantum Fisher information from dynamic susceptibilities \cite{haukeetal2016}, extracted from neutron scattering data in Heisenberg chains \cite{mathewetal_2020,scheieetal_2021}. In particular, multipartite entanglement was recently detected in the quantum-critical regime of a triangular-lattice antiferromagnet \cite{scheieetal_QCP}.

\subsection{Other systems}
Besides the platforms discussed so far, several other systems have demonstrated their potential for quantum technologies. Examples include defects in diamond, quantum dots nuclear magnetic resonance (NMR) processors, silicon spin qubits, and solid-state mechanical oscillators, just to name a few. Using these as building blocks for multi-partite quantum systems is still in a preliminary stage, as significant efforts are still being put in developing a toolbox for their operation, or in engineering couplings between them. For example, entanglement has so far been demonstrated for three spins in diamond \cite{doi:10.1126/science.1157233,Waldherr_2014}, and for three spins in silicon \cite{Takeda_2021}.

\section{Conclusions}\label{sec5}

\subsection{Open problems}
We conclude this review by mentioning a few relevant open questions and research directions for future works, although the proposed list is far from being comprehensive. \\

\noindent\textit{Entanglement quantification.--} Beyond entanglement detection, the problem of quantifying entanglement in multipartite systems in a scalable way remains largely open, although first studies have been published \cite{FadelPRL21}. While relevant concepts for entanglement quantification have been reviewed in ref.~\cite{friietal_review_2018}, their implementation to shed light \eg on problems such as entanglement dynamics after a quantum quench, is an open issue. \\

\noindent\textit{Probing entanglement in topological phases.--} While topological quantum many-body phases are defined by nonlocal order parameters, which manifest themselves through characteristic signatures in entanglement entropies, finding signatures of interacting topological phases through tailored entanglement witnesses, or even Bell's inequalities, represents a very intriguing perspective.\\

\noindent\textit{Continuous-variable systems.--} The data-driven approaches presented here are intrinsically tailored to discrete-variable systems, such as qubits, or of qudits of modest local Hilbert space dimension. Systematic extensions to continuous-variable systems, describing \eg quadratures of bosonic modes, would be of great interest. \\

\noindent\textit{Bell inequalities for multipartite non-Gaussian states.--} Inequalities involving permutationally-invariant one- and two-body correlators were proven successful in detecting Bell correlations in spin-squeezed atomic ensembles. However, these are intrinsically limited to Gaussian states, as more general states require high-order correlators to be characterized. Of great experimental interest would be to find Bell correlation witnesses for non-Gaussian spin states, that can be probed through collective measurements.\\

\noindent\textit{Bipartite coarse-grained Bell inequalities.--} Although multipartite Bell inequalities for many-body systems have been derived and (witnesses thereof) violated in experiments, it remains a major open problem to find bipartite Bell inequalities violated in a scalable way by many-body system. As a concrete example, on may consider the detection of nonlocality between two partitions of a spin-squeezed atomic ensemble on which collective spin measurements are performed \cite{Jing2019,KitzingerPRA21}.\\

\noindent\textit{Role of Bell nonlocality in metrology.--} Although it is known that Bell nonlocal states could result in a large quantum Fisher information \cite{FrowisDoes19}, it is not clear whether they enable metrological tasks for which entanglement alone would be insufficient. In the case of EPR steering a similar question has been answer affirmatively in ref.~\cite{YadinNatCom21}, and it would be natural to aim at extending such results to nonlocality.\\

\noindent\textit{Structure of multipartite nonlocality.--} Witnesses quantifying $k$-producibility of Bell correlation for $k<7$ have been derived, and tested experimentally. For larger $k$ the approach adopted in ref.~\cite{BaccariDepth} was not scalable, and alternatives to it are not clear. A breakthrough would be to find a Bell correlation witness involving collective spin measurements, capable of quantifying arbitrary $k$-producibility.\\

\noindent\textit{Multipartite self-testing.--} The concept of self-testing allows one to draw conclusions on the system dimensionality, prepared state, and performed measurements, relying only on the observed statistics in a device-independent scenario. While this has been well analyzed in the bipartite scenario \cite{Supic2020selftestingof}, it is not clear whether it would be possible to formulate a scalable approach for self-testing in multipartite scenarios. First studies in this direction have focused on multiqubit graph states \cite{baccarietal2020} and many-body spin singlets \cite{frerotA2021}. \\

\noindent\textit{Network nonlocality.--} A new form of Bell nonlocality has been recently investigated in networks (see \cite{Tavakoli_2022} for a review), and it represents a more general paradigm than the standard Bell test as considered in this review. Applying such concepts to characterize many-body entanglement from a renewed perspective is a completely open field of research, that could turn out extremely useful in the context of quantum communication.\\

\noindent\textit{Multipartite EPR steering.--} While seminal works on multipartite EPR steering have focused on typically three subsystems, systematic studies in the multipartite scenario have not been reported. Promising directions include the derivation of inequalities based on covariances among subsystems to probe genuine multipartite steering in a collection of photonic modes \cite{tehetal2022}. Moreover, a complete understanding of monogamy relations is still lacking.

\subsection{Summary of the review}
The phenomenon of quantum entanglement is among the most intriguing manifestations of quantum mechanics in composite systems. Its various manifestations, from the violation of a naive notion of locality (through the violation of Bell inequalities), to the apparent possibility to manipulate remotely the state of a quantum system entangled with another one (through the phenomenon of EPR steering), depart from any classical intuition and challenge our understanding of nature. Despite, and probably because of this, quantum correlations turned out to be irreplaceable resources for a variety of tasks such as quantum metrology, quantum communications and quantum computing, that would be inaccessible by classical means. This has triggered an immense interest in various research communities towards finding useful practical applications for quantum technologies.

In recent years, spectacular progress has been made in the manipulation of entangled quantum many-body states with a variety of experimental platforms. This motivated a large body of works, both theoretical and experimental, to further understand and characterize the whole palette of correlations that can appear in multipartite scenarios. 

On the one hand, we have reviewed recent theoretical developments allowing one to characterize entanglement, EPR steering and Bell nonlocality from experimental data, as the one obtained from existing or near-term quantum devices. We have emphasized the intrinsic challenges associated to the exponential growth in complexity of many-body systems, and how these could be circumvented. On the other hand, we have reviewed seminal experimental works demonstrating the preparation, control, and measurements of quantum many-body states in the laboratory. We put particular focus on describing the approaches used to prepare highly entangled states, and the tools available to probe them. Finally, we have presented a list of relevant open problems and research directions.

Although the progress achieved in the field over recent years has been spectacular, a vast number of conceptual questions and experimental challenges still remain.
As it should be clear from this review, the quest for understanding and characterising multipartite quantum correlations is just at the beginning.

\subsection{Acknowledgements}
The three authors are grateful to an entire community of colleagues working tirelessly on expanding our understanding of many-body systems, from foundational aspects to technological applications. 
We have benefited, and continue to benefit, from countless discussions with so many colleagues that would be impossible to acknowledge all of them here individually. We have not doubt that they will recognize their contribution and our deepest gratitude.\\
\noindent I.F. acknowledges support from the European Union’s Horizon 2020 research and innovation programme under the Marie-Skłodowska-Curie grant agreement No 101031549 (QuoMoDys).\\
M.F. was supported by The Branco Weiss Fellowship -- Society in Science, administered by the ETH Z\"{u}rich.\\
M.L. acknowledges support from: ERC AdG NOQIA; Ministerio de Ciencia y Innovation Agencia Estatal de Investigaciones (PGC2018-097027-B-I00/10.13039/501100011033,  CEX2019-000910-S/10.13039/501100011033, Plan National FIDEUA PID2019-106901GB-I00, FPI, QUANTERA MAQS PCI2019-111828-2, QUANTERA DYNAMITE PCI2022-132919,  Proyectos de I+D+I “Retos Colaboración” QUSPIN RTC2019-007196-7); MICIIN with funding from European Union NextGenerationEU (PRTR-C17.I1) and by Generalitat de Catalunya;  Fundació Cellex; Fundació Mir-Puig; Generalitat de Catalunya (European Social Fund FEDER and CERCA program, AGAUR Grant No. 2021 SGR 01452, QuantumCAT \ U16-011424, co-funded by ERDF Operational Program of Catalonia 2014-2020); Barcelona Supercomputing Center MareNostrum (FI-2022-1-0042); EU Horizon 2020 FET-OPEN OPTOlogic (Grant No 899794); EU Horizon Europe Program (Grant Agreement 101080086 — NeQST), National Science Centre, Poland (Symfonia Grant No. 2016/20/W/ST4/00314); ICFO Internal “QuantumGaudi” project; European Union’s Horizon 2020 research and innovation program under the Marie-Skłodowska-Curie grant agreement No 101029393 (STREDCH) and No 847648  (“La Caixa” Junior Leaders fellowships ID100010434: LCF/BQ/PI19/11690013, LCF/BQ/PI20/11760031,  LCF/BQ/PR20/11770012, LCF/BQ/PR21/11840013). Views and opinions expressed in this work are, however, those of the author(s) only and do not necessarily reflect those of the European Union, European Climate, Infrastructure and Environment Executive Agency (CINEA), nor any other granting authority.  Neither the European Union nor any granting authority can be held responsible for them.

\bibliography{biblio}

\end{document}